\documentclass[twocolumn,preprintnumbers,amsmath,amssymb,superscriptaddress,nofootinbib,longbibliography]{revtex4-1}

\usepackage{graphicx}
\usepackage{bm}
\usepackage{dsfont}
\usepackage[usenames,dvipsnames]{xcolor}
\usepackage{pstricks}
\usepackage[tight]{subfigure}
\usepackage{verbatim}
\usepackage{units}
\usepackage{multirow}
\usepackage{enumitem}
\usepackage{mathrsfs}
\usepackage{leftidx}
\usepackage{xspace}

\usepackage[a4paper]{hyperref}
\hypersetup{colorlinks=true,linktoc=all,linkcolor=blue,breaklinks=true,citecolor=blue,urlcolor=blue}

\usepackage[english]{babel}
\newcommand{\Ket}[1]{\big|#1\big\rangle}

\newcommand{\GF}[2]{\langle\langle #1|#2\rangle\rangle}
\newcommand{\acG}{\mathcal{G}}
\newcommand{\sfA}{\mathcal{A}}

\newcommand{\TK}{T_\text{K}}
\newcommand{\TKqd}{T_\text{K}^0}
\newcommand{\Ds}{D_\text{spin}}

\newcommand{\GAP}{G_\text{AP}}

\newcommand{\Ham}{\skew{3}{\hat}{\mathcal{H}}}
\newcommand{\opS}{\skew{3}{\hat}{S}}
\newcommand{\opc}{\skew{2}{\hat}{c}}
\newcommand{\opn}{\skew{2}{\hat}{n}}
\newcommand{\opa}{\skew{1}{\hat}{a}}
\newcommand{\opf}{\skew{3}{\hat}{f}}
\newcommand{\opQ}{\skew{2}{\hat}{Q}}
\newcommand{\opI}{\skew{3}{\hat}{I}}
\newcommand{\opII}{\skew{3}{\hat}{\mathcal{I}}}
\newcommand{\oper}[1]{\skew{3}{\hat}{#1}}
\newcommand{\opx}{\skew{1}{\hat}{x}}

\newcommand{\intd}{\text{d}}
\newcommand{\via}{\emph{via} }

\begin{document}


\title{Spin-resolved dynamical conductance of correlated large-spin magnetic molecules}

\author{Anna P\l omi\'nska}
\email{anna.plominska@amu.edu.pl}
\affiliation{Faculty of Physics, Adam Mickiewicz University, 61-614 Pozna\'{n}, Poland}
\author{Maciej Misiorny}
\email{misiorny@amu.edu.pl}
\affiliation{Department of Microtechnology and Nanoscience MC2, Chalmers University of Technology, SE-412 96 G\"{o}teborg, Sweden}
\affiliation{Faculty of Physics, Adam Mickiewicz University, 61-614 Pozna\'{n}, Poland}
\author{Ireneusz Weymann}
\affiliation{Faculty of Physics, Adam Mickiewicz University, 61-614 Pozna\'{n}, Poland}

\date{\today}

\begin{abstract}
The finite-frequency transport properties of a large-spin molecule
attached to ferromagnetic contacts are studied theoretically in the Kondo regime.
The focus is on the behavior of the dynamical conductance in the linear response regime,
which is determined by using the numerical renormalization group method.
It is shown that the dynamical conductance depends greatly on the magnetic configuration
of the device and intrinsic parameters of the molecule.
In particular, conductance exhibits characteristic features for frequencies
corresponding to the dipolar and quadrupolar exchange fields
resulting from the presence of spin-dependent tunneling.
Moreover, a dynamical spin accumulation in the molecule,
associated with the off-diagonal-in-spin component of the conductance, is predicted.
This spin accumulation becomes enhanced with increasing the 
spin polarization of the leads, and it results in a nonmonotonic dependence
of the conductance on frequency, with local maxima
occurring for characteristic energy scales.
\end{abstract}

\maketitle

\section{Introduction}

Magnetic molecules are envisaged as a prospective base for future
storage and information processing at the nanoscale
\cite{Kahn_Science279/1998,Leuenberger_Nature410/2001,
Rocha_NatureMater.4/2005,Mannini_NatureMater.8/2009,Bartolome_book}.
To realize this dream, it is, however, necessary
to provide a detailed understanding of various properties
and mechanisms governing transport properties
of molecules attached to external leads.
The steady-state transport behavior of
molecules coupled to metallic electrodes
has already been a subject of extensive investigations
\cite{Sasaki_Nature405/2000,Nygaard_Nature(London)408/2000,Liang_Nature417/2002,
Zhao_Science309/2005,Romeike_Phys.Rev.Lett.96/2006,Romeike_Phys.Rev.Lett.97/2006,
Otte_NaturePhys.4/2008,Roch_Nature453/2008,Scott_ACSNano4/2010,Parks_Science328/2010}.
First of all, it turns out that when the coupling to external leads
is relatively weak transport is dominated by single-electron charging effects
\cite{Grabert_/1992}.
On the other hand, in the strong coupling regime the many-body correlations
can result in the Kondo effect
\cite{Kondo_Prog.Theor.Phys32/1964,Goldhaber_Nature391/98,Cronenwett_Science281/1998}.
For molecules with spin one-half, this phenomenon manifests as an enhancement of linear conductance 
for temperatures lower than some characteristic temperature \mbox{---the} so-called Kondo temperature $\TK$~\cite{Hewson_book}. Interestingly, for molecules exhibiting spin \emph{larger} than one-half more exotic Kondo effects can in general emerge, whose typical examples are the underscreened \cite{Koller_Phys.Rev.B72/2005,Roch_Phys.Rev.Lett.103/2009,
Weymann_Phys.Rev.B81/2010,Cornaglia_Europhys.Lett.93/2011,
Misiorny_Phys.Rev.B86/2012_UK}
and the two-stage Kondo phenomena
\cite{Wiel_Phys.Rev.Lett.88/2002,Posazhennikova_Phys.Rev.B75/2007,
Zitko_J.Phys.:Condens.Matter22/2010,Misiorny_Phys.Rev.B86/2012,Wojcik_PhysRevB.91.134422/2014}.
In the former case, the low-temperature conductance still
approaches the conductance quantum, while in the latter case,
a suppression of transport takes place.
Moreover, the stationary transport through magnetic molecules
has been also considered in the case of ferromagnetic leads
\cite{Misiorny_Phys.Rev.Lett.106/2011,Misiorny_Phys.Rev.B84/2011,
Misiorny_Phys.Rev.B86/2012,Misiorny_Phys.Rev.B90/2014,Misiorny_NaturePhys.9/2013}.
In the strong coupling regime, it was shown that 
the occurrence of the Kondo effect is conditioned by the presence
of a dipolar exchange  field~\cite{Martinek_Phys.Rev.Lett.91/2003_127203},
which effectively acts as an external magnetic field. If such a field exceeds $\TK$,
the Kondo phenomenon becomes suppressed
\cite{Misiorny_Phys.Rev.Lett.106/2011,Misiorny_Phys.Rev.B84/2011}.
Importantly, for molecules with spin larger than one-half, an additional
quadrupolar field arises~\cite{Misiorny_NaturePhys.9/2013}.
Since this field essentially imposes anisotropy on the molecular spin, it can thus 
also destroy the Kondo effect once larger than the Kondo temperature.

In the present paper we focus on on the \emph{frequency-dependent} transport through large-spin molecules
coupled to ferromagnetic contacts in the Kondo regime. The motivation for this study stems from the fact that, unlike for the stationary transport, the analysis of dynamical response of a system to  an applied time-dependent bias 
provides an additional insight into the fluctuations in the system~\cite{Heikkilae_book}.
In particular, the spin-dependent component of the dynamical conductance 
$\acG_{\sigma\sigma^\prime}^{qq^\prime}(\omega)$,
describing the current response in the $q$th electrode to all external voltages
applied to $q^\prime$th electrodes
\begin{equation}
	I^q(\omega)
	=
	\sum_{q^\prime\sigma\sigma^\prime}
	\acG_{\sigma\sigma^\prime}^{qq^\prime}(\omega)
	V^{q^\prime}(\omega)
	,
\end{equation}
is related to the corresponding component of the symmetrized-noise power spectral density, 
\begin{equation}
	[\mathcal{S}_\text{I}]_{\sigma\sigma^\prime}^{qq^\prime}(\omega)
	=	
	\frac{1}{2}
	\int\!\!\intd\tau\,
	\text{e}^{-i\omega\tau}
	\big\langle
	\big\{
		\opI_\sigma^q(\tau),
		\opI_{\sigma^\prime}^{q^\prime}(0)
	\big\}
	\big\rangle
	,
\end{equation}
through the \emph{fluctuation-dissipation theorem}~\cite{Kubo_book}
\begin{equation}
	[\mathcal{S}_\text{I}]_{\sigma\sigma^\prime}^{qq^\prime}(\omega)
	=
	-\hbar\omega
	\coth\Big(\frac{\hbar\omega}{2k_\text{B}T}\Big)
	\acG_{\sigma\sigma^\prime}^{qq^\prime}(\omega)
	.
\end{equation}
Thus, at very low temperatures, as considered in this paper,
the noise normalized to frequency can be directly accessed from the dynamical conductance.

We note that the dynamical aspect of spin-dependent transport in the Kondo regime has so far only been addressed theoretically in the case of quantum dots
\cite{Sindel_Phys.Rev.Lett.94/2005,Toth_Phys.Rev.B76/2007,
Moca_Phys.Rev.B81/2010,Moca_Phys.Rev.B84/2011,
Moca_Phys.Rev.B83/2011,Weymann_J.Appl.Phys.109/2011,Moca_Phys.Rev.B89/2014},
but not in magnetic molecules. Specifically, it was shown that finite-frequency conductance
provides a direct access to the equilibrium spectral function of the system and the Kondo resonance,
which otherwise would have to be probed under nonequilibrium conditions induced by application of a finite bias voltage \cite{Sindel_Phys.Rev.Lett.94/2005}. Furthermore, the analysis of the spin-resolved conductance for the Kondo model,
revealed the effect of a dynamical spin accumulation,
which is related to finite off-diagonal component of conductance in the spin space
\cite{Moca_Phys.Rev.B81/2010,Moca_Phys.Rev.B84/2011}.
It is also important to notice that the
finite-frequency transport properties of quantum dot systems
have also been explored experimentally in a wide range of frequencies
\cite{Zakka_PRL.99/2007,Gabelli_PRL.100/2008,
Zakka_PRL.104/206802,Basset_PRL.105/166801,Basset_PRL.108/046802}.

Here, we investigate the behavior of the frequency-dependent conductance
for a magnetic molecule, modeled as a large-spin magnetic core exchange-coupled to a single conducting orbital level ---that is, only the orbital level is assumed to be tunnel-coupled to external ferromagnetic leads. The analysis is
performed in the linear response regime by using the Kubo formula,
and the relevant correlation functions are determined with the aid of
the numerical renormalization group (NRG) method
\cite{Wilson_RMP.47/773}.
We show that the dynamical conductance strongly depends
on the magnetic configuration of the device
and intrinsic parameters of the molecule, such as magnetic anisotropy
and exchange coupling between the orbital level and magnetic core.
When spin moments of leads are oriented in  parallel,
the presence of dipolar and quadrupolar exchange fields
suppresses the conductance. On the other hand,
for the antiparallel configuration we find a nontrivial behavior
of the conductance on frequency depending on the type
of exchange interaction. Furthermore,
we predict a dynamical spin accumulation in the molecule
triggered by time-dependent bias, which
strongly depends on intrinsic parameters of the molecule
and in general exhibits a nonmonotonic dependence on frequency.

The paper is organized as follows: In Sec.~\ref{sec:Model}
the model and key Hamiltonians for the considered system are introduced.
Next, a detailed derivation of the formula for the dynamic conductance is presented in Sec.~\ref{sec:Dyn_cond}.
The numerical renormalization group method used for calculations
is briefly described in Sec.~\ref{sec:NRG}.
Numerical results are discussed in Secs.~\ref{sec:QD} and~\ref{sec:Large_spin}.
In particular, two distinctive cases are analyzed:
for a single-level quantum dot of spin $S=1/2$ (Sec.~\ref{sec:QD}),
and for a large-spin ($S>1/2$) molecule that can additionally exhibit
uniaxial magnetic anisotropy (Sec.~\ref{sec:Large_spin}).
Finally, the summary and key conclusions
of the paper are given in Sec.~\ref{sec:Conclusions}.

\section{Theoretical description}

\subsection{\label{sec:Model}Model}
%
\begin{figure}[t]
	\includegraphics[width=0.75\columnwidth]{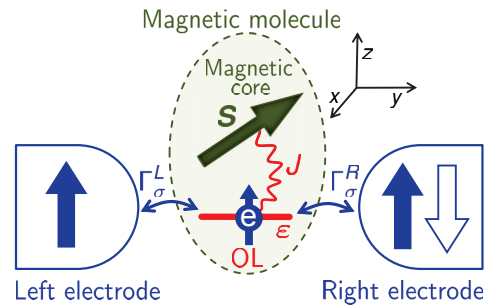}
	\caption{
		(color online)
		Schematic representation of the model system under consideration ---a large-spin molecule embedded in a magnetic tunnel junction. For detailed description see Sec.~\ref{sec:Model}.
		\label{fig1}}
\end{figure}

We consider a model of a large-spin molecule embedded in a magnetic tunnel junction, as schematically depicted in Fig.~\ref{fig1}. The molecule, which in the following will be also referred to as a magnetic quantum dot (MQD), is assumed here to consist of a single conducting orbital level (OL) and a magnetic core (internal spin) represented by a spin operator~$\oper{\bm{S}}$. Moreover, tunnel coupling of the OL to two metallic ferromagnetic electrodes of the junction, whose strength is described by  spin-dependent hybridization functions $\Gamma_\sigma^q$ [$q=L(\text{left}),R(\text{right})$], enables transport of electrons across the junction \via the molecule. Importantly, when the OL is occupied by a single electron its spin becomes coupled \via exchange interaction~$J$ to the spin of the magnetic core.
The model setup to be studied is thus characterized by the total Hamiltonian of the general form
\begin{equation}\label{eq:H_total}
	\Ham
	=
	\Ham_\textrm{MQD} + \Ham_\textrm{el} + \Ham_\textrm{tun}
	,
\end{equation}
where the three terms represent: the molecule, $\Ham_\textrm{MQD}$, electrodes, $\Ham_\textrm{el}$, and electron tunneling processes, $\Ham_\textrm{tun}$.

%
%
To begin with, the first term of the total Hamiltonian~(\ref{eq:H_total}), standing for the MQD, in the absence of an external magnetic field generally comprises the following parts
\begin{equation}\label{eq:H_MQD}
	\Ham_\textrm{MQD}
	=
	\Ham_\textrm{MA}+\Ham_\textrm{OL}+\Ham_\textrm{J}.
\end{equation}
Here, of key importance is the first part,
\begin{equation}\label{eq:H_MA}
	\Ham_\textrm{MA}
	=
	-D\opS_z^2
	,
\end{equation}
describing the lowest order magnetic anisotropy of the MQD's magnetic core, with $D$ denoting the \emph{uniaxial} anisotropy constant.
Next, the essential features of the conducting orbital are captured by the second term of the Hamiltonian~(\ref{eq:H_MQD}),
\begin{equation}
	\Ham_\textrm{OL}
	=
	\varepsilon\sum_\sigma\opn_\sigma
	+
	U\opn_\uparrow \opn_\downarrow
	,
\end{equation}
where $\opn_\sigma=\opc_\sigma^\dagger \opc_\sigma^{}$ and the operator $\opc_\sigma^\dagger\  (\opc_\sigma^{})$ is responsible for creation (annihilation) of an electron with spin $\sigma$ and energy $\varepsilon$ in the OL. Note that if the orbital is simultaneously occupied by two electrons of opposite spin this results in occurrence of the Coulomb energy~$U$.
Finally, the exchange interaction between the spin of a single electron occupying the OL,
$
	\oper{\bm{s}}
	=
	(1/2)\sum_{\sigma\sigma^\prime}
	\hat{\bm{\sigma}}_{\sigma\sigma^\prime}
	\opc_\sigma^\dagger\opc_{\sigma^\prime}^{}
$
[with $\hat{\bm{\sigma}}\equiv(\hat{\sigma}_x,\hat{\sigma}_y,\hat{\sigma}_z)$ denoting the Pauli spin operator], and the magnetic core,~$\oper{\bm{S}}$, is represented by the third term,
\begin{equation}\label{eq:H_J}
	\Ham_\text{J}
	=
	-J\oper{\bm{s}}\cdot\oper{\bm{S}}
	,
\end{equation}
and it will be referred to as \emph{ferromagnetic} (FM) for $J>0$ and \emph{antiferromagnetic} (AFM) for $J<0$. Importantly, if $J=0$, the model under discussion reduces to a single-level quantum dot~(QD), with only the orbital part $\Ham_\textrm{OL}$, corresponding to the Anderson single impurity model, being the only relevant term.

%
%
Electrodes of the magnetic tunnel junction are represented as reservoirs of spin-polarized and non-interacting itinerant electrons, so that, with creation [annihilation] of a spin-$\sigma$ electron in the $q$th electrode described by the operator $\opa_\sigma^{q\dagger}(\epsilon)\ \big[\opa_\sigma^{q}(\epsilon)\big]$, the second term of the total Hamiltonian~(\ref{eq:H_total}) is given by
\begin{equation}\label{eq:H_el}
	\Ham_\text{el}
	=
	\sum_{q\sigma}
	\int_{-W}^W
	\intd\epsilon\,
	\epsilon\,
	\opa_\sigma^{q\dag}\!(\epsilon)
	\opa_\sigma^q(\epsilon)
	,
\end{equation}
with $W$ denoting the conduction band half-width. Moreover, we assume that spin moments of electrodes form a collinear configuration, that is, they can be oriented with respect to each other either parallel (P) or antiparallel (AP),
see Fig.~\ref{fig1}.
The assumption is also made that these spin moments are collinear with the principal magnetic ($z$) axis of the MQD set by the uniaxial component of magnetic anisotropy.
In consequence, the process of electron tunneling across the junction \via the OL of the molecule, taken into account by the final term of the total Hamiltonian~(\ref{eq:H_total}), is expressed as
\begin{equation}\label{eq:H_tun}
	\Ham_\textrm{tun}
	=
	\sum_{q\sigma}
	\sqrt{\frac{\Gamma_\sigma^q}{\pi}}
	\int_{-W}^W
	\intd\epsilon\,
	\Big[
	\opa_\sigma^{q\dagger}\!(\epsilon) \opc_\sigma^{}
	+
	\opc_\sigma^\dagger \opa_\sigma^q\!(\epsilon)
	\Big]
	.
\end{equation}
The spin-dependent hybridization functions $\Gamma_\sigma^q$ in the equation above can be further conveniently parametrized by means of the \emph{spin-full hybridization} (OL broadening) due to tunnel coupling to the $q$th electrode,
$
	\Gamma^q
	=
	\Gamma_\uparrow^q+\Gamma_\downarrow^q
$,
and the \emph{spin polarization coefficient} $p^q$ defined as
$
	p^q
	=
	\big(\Gamma_\uparrow^q-\Gamma_\downarrow^q\big)
	/
	\big(\Gamma_\uparrow^q+\Gamma_\downarrow^q\big)
$.
Assuming identical electrodes and tunnel barriers, that is, $\Gamma^L=\Gamma^R=\Gamma$ and $p^L=p^R=p$, one obtains $\Gamma^L_{\uparrow(\downarrow)}=\Gamma^R_{\uparrow(\downarrow)}=(\Gamma/2)(1\pm p)$ for the parallel magnetic configuration, and $\Gamma^L_{\uparrow(\downarrow)}=\Gamma^R_{\downarrow(\uparrow)}=(\Gamma/2)(1\pm p)$ for the antiparallel one.

However, it has been demonstrated that for calculations within the linear response theory
it becomes advantageous to use a canonical transformation~\cite{Glazman_JETP.Lett.47/1988,Bruus_book,Misiorny_Phys.Rev.B84/2011} corresponding to a rotation in the space of left-right electron operators that allows for decoupling of the OL from the \emph{odd} linear combination of electrode operators,
$
	\opa_\sigma^o(\epsilon)
	=
	\Lambda_\sigma^R
	\opa_\sigma^L(\epsilon)
	-
	\Lambda_\sigma^L
	\opa_\sigma^R(\epsilon)
$
with $\Lambda_\sigma^q=\sqrt{\Gamma_\sigma^q/(\Gamma_\sigma^L+\Gamma_\sigma^R)}$. The orbital couples then only to a single effective electron reservoir constructed out of the \emph{even} linear combinations of electrode operators,
$
	\opa_\sigma^e(\epsilon)
	=
	\Lambda_\sigma^L
	\opa_\sigma^L(\epsilon)
	+
	\Lambda_\sigma^R
	\opa_\sigma^R(\epsilon)
$.
As a result, in the new basis of even-odd electron operators the tunneling Hamiltonian~(\ref{eq:H_tun}) is transformed as follows
\begin{equation}\label{eq:H_tun_eff}
	\Ham_\textrm{tun}
	=
	\sum_{q\sigma}
	\sqrt{\frac{\Gamma_\sigma^\text{eff}}{\pi}}
	\int_{-W}^W
	\intd\epsilon\,
	\Big[
	\opa_\sigma^{e\dagger}\!(\epsilon) \opc_\sigma^{}
	+
	\opc_\sigma^\dagger \opa_\sigma^e\!(\epsilon)
	\Big],
\end{equation}
with $\Gamma_\sigma^\text{eff}=\Gamma_\sigma^L+\Gamma_\sigma^R$. Importantly, note that in the antiparallel magnetic configuration $\Gamma_\sigma^\text{eff}=\Gamma$, whereas in the parallel one $\Gamma_{\uparrow(\downarrow)}^\text{eff}=\Gamma(1\pm p)$.

\subsection{\label{sec:Dyn_cond}Dynamical conductance}

To study the dynamical response of the system we apply an external time-dependent bias voltage $V^{L(R)}(t)$, 
which is described by a new term $\Ham_\text{bias}$~\cite{Toth_Phys.Rev.B76/2007,Weymann_J.Appl.Phys.109/2011},
\begin{equation}
	\Ham_\text{bias}
	=
	\sum_q
	\opQ^q
	V^q(t)
	,
\end{equation}
added to the total Hamiltonian~(\ref{eq:H_total}),
with $\opQ^q=\sum_\sigma\opQ_\sigma^q$, and
\begin{equation}\label{eq:charge_op}
	\opQ_\sigma^q
	=
	-|e|
	\int_{-W}^W\intd\epsilon\,
	\opa_\sigma^{q\dagger}(\epsilon)
	\opa_\sigma^{q}(\epsilon)
\end{equation}
denoting the operator for the spin-$\sigma$ component of charge induced in the $q$th electrode.
Consequently, in the linear response, the current flowing through the MQD
can be described by the Kubo formula
\begin{equation}\label{eq:Kubo_formula}
	\langle \opI^q(t)\rangle
	=
	\sum_{q^\prime\sigma\sigma^\prime}
	\int\!\!\textrm{d}t^\prime\,
	\acG_{\sigma\sigma^\prime}^{qq^\prime}(t-t^\prime)
	V^{q^\prime}(t^\prime),
\end{equation}
where $\acG_{\sigma\sigma^\prime}^{qq^\prime}(t-t^\prime)$
is the time-dependent response of the system (conductance), 
induced by applied time-dependent voltage, 
given explicitly by
\begin{equation}\label{eq:G_vs_GF}
	\acG_{\sigma\sigma^\prime}^{qq^\prime}(t-t^\prime)
	=
	-\frac{i}{\hbar}\theta(t-t^\prime)
	\big\langle[
	\opI_\sigma^q(t),
	\opQ_{\sigma^\prime}^{q^\prime}(t^\prime)
	]\big\rangle,
\end{equation}
with $\opI_\sigma^q(t)=\text{d}\opQ_\sigma^q(t)/\text{d}t$ and $\opQ_{\sigma^\prime}^{q^\prime}(t^\prime)$, Eq.~(\ref{eq:charge_op}), standing for the current and charge operator in the interaction picture, respectively, and $\langle\ldots\rangle$ denoting the quantum-statistical averaging.
Using the above expression, the frequency-dependent conductance (admittance) $\acG_{\sigma\sigma^\prime}^{qq^\prime}(\omega)$ can be shown to take the form (for details see App.~\ref{app:aux_deriv})
\begin{equation}\label{eq:G_omega_def}
	\acG_{\sigma\sigma^\prime}^{qq^\prime}(\omega)
	=
	\frac{i}{\omega}
	\Big[\GF{\opI_\sigma^q}{\opI_{\sigma^\prime}^{q^\prime}}_\omega^\textrm{r}
	-
	\GF{\opI_\sigma^q}{\opI_{\sigma^\prime}^{q^\prime}}_{\omega=0}^\textrm{r}\Big].
\end{equation}
In order to find the frequency-dependent conductance  $\acG_{\sigma\sigma^\prime}^{qq^\prime}(\omega)$, one has to calculate $\GF{\opI_\sigma^q}{\opI_{\sigma^\prime}^{q^\prime}}_\omega^\textrm{r}$, that is, the Fourier transform of the retarded Green's function for the current operator defined as
$
	\GF{\opI_\sigma^q}{\opI_{\sigma^\prime}^{q^\prime}}_t^\textrm{r}
	=
	-(i/\hbar)\theta(t)
	\big\langle[
	\opI_\sigma^q(t),
	\opI_{\sigma^\prime}^{q^\prime}(0)
	]\big\rangle
$.
Although in a general case, derivation of such a function can pose a serious challenge, in the situation under consideration the MQD is coupled only to the even reservoir, see Sec.~\ref{sec:Model}, which will lead to a great simplification of analytical formulae.

It turns out that due to application of the canonical transformation discussed above, also the current operator~$\opI_\sigma^q$ can  be separated into two parts representing even~($\opI_\sigma^{q,e}$) and odd~($\opI_\sigma^{q,o}$) transport channel, $\opI_\sigma^q=\opI_\sigma^{q,e}+\opI_\sigma^{q,o}$, with
\begin{equation}\label{eq:opIeo_def}
	\left\{
	\begin{aligned}
	&\opI_\sigma^{q,e}
	=
	-i\frac{|e|}{\hbar\sqrt{\pi\rho}}
	\cdot
	\frac{\Gamma_\sigma^q}{\sqrt{\Gamma_\sigma^\text{eff}}}
	\cdot
	\opII_\sigma^e	
	,
	\\
	&\opI_\sigma^{q,o}
	=
	-i\eta_q\frac{|e|}{\hbar\sqrt{\pi\rho}}
	\cdot
	\sqrt{
	\frac{\Gamma_\sigma^L\Gamma_\sigma^R}{\Gamma_\sigma^\text{eff}}	
	}
	\cdot
	\opII_\sigma^o	
	.
	\end{aligned}
	\right.
\end{equation}
Here, the coefficient $\eta_q$ should be understood as $\eta_L=1$ and $\eta_R=-1$, whereas $\opII_\sigma^{e(o)}$ denotes the dimensionless current operator given by
\begin{equation}\label{eq:opIIeo_def}
	\opII_\sigma^{e(o)}
	=
	\opc_\sigma^\dagger \hat{\Psi}_\sigma^{e(o)}
	-
	\hat{\Psi}_\sigma^{e(o)\dagger}\opc_\sigma^{}
	,
\end{equation}
with $\hat{\Psi}_\sigma^{e(o)}=\sqrt{\rho}\int\intd\epsilon\,\opa_\sigma^{e(o)}(\epsilon)$ and $\rho=1/(2W)$
being the density of states of a conduction band.
Furthermore, employing the above decomposition of the current operator~$\opI_\sigma^q$,
the Green's function~$\GF{\opI_\sigma^q}{\opI_{\sigma^\prime}^{q^\prime}}_\omega^\textrm{r}$  occurring in the expression for dynamical conductance $G_{\sigma\sigma^\prime}^{qq^\prime}(\omega)$, Eq.~(\ref{eq:G_omega_def}), can be split into two components corresponding to the even and odd channels,
\begin{equation}\label{eq:GF_odd_even}
	\GF{\opI_\sigma^q}{\opI_{\sigma^\prime}^{q^\prime}}_\omega^\textrm{r}
	=
	\GF{\opI_\sigma^{q,e}}{\opI_{\sigma^\prime}^{q^\prime\!,e}}_\omega^\textrm{r}
	+
	\delta_{\sigma\sigma^\prime}
	\GF{\opI_\sigma^{q,o}}{\opI_{\sigma^\prime}^{q^\prime\!,o}}_\omega^\textrm{r}
	.
\end{equation}
The explicit expressions for these two functions are found to be of the following form (for details of derivation see App.~\ref{app:aux_deriv})
\begin{equation}\label{eq:GF_even}
	\GF{\opI_\sigma^{q,e}}{\opI_{\sigma^\prime}^{q^\prime\!,e}}_\omega^\textrm{r}
	=
	-\frac{G_0}{\hbar\rho}
	\cdot
	\frac{\Gamma_\sigma^q\Gamma_{\sigma^\prime}^{q^\prime}}
	{\sqrt{\Gamma_\sigma^\text{eff}\Gamma_{\sigma^\prime}^\text{eff}}}
	\GF{\opII_\sigma^e}{\opII_{\sigma^\prime}^{e\dagger}}_\omega^\textrm{r},
\end{equation}
and
\begin{align}\label{eq:GF_odd}
	&\GF{\opI_\sigma^{q,o}}{\opI_{\sigma}^{q^\prime\!,o}}_\omega^\textrm{r} =
	\eta_q\eta_{q^\prime}
	G_0\cdot
	\frac{\Gamma_{\sigma}^L\Gamma_{\sigma}^R}{\Gamma_{\sigma}^\text{eff}}
	\nonumber\\
	&\hspace*{2pt}\times\!\int\textrm{d}\omega^\prime \Big\{
	\big[f(\omega^\prime-\omega)+f(\omega^\prime+\omega)\big]
	\textrm{Re}\GF{\opc_\sigma^{}}{\opc_\sigma^\dagger}_{\omega^\prime}^\textrm{r}
	\nonumber\\
	&\hspace*{31pt}+
	i\big[f(\omega^\prime-\omega)-f(\omega^\prime+\omega)\big]
	\textrm{Im}\GF{\opc_\sigma^{}}{\opc_\sigma^\dagger}_{\omega^\prime}^\textrm{r}
	\Big\}
	,
	\!\!
\end{align}
with $G_0\equiv2e^2/h$. Additionally, in the equation above $\GF{\opc_\sigma^{}}{\opc_\sigma^\dagger}_{\omega}^\textrm{r}$ represents the Fourier transform of the retarded Green's function for the OL, and $f(\omega)=\big\{1+\exp[\hbar\omega/(k_\text{B}T)]\big\}^{-1}$ denotes the Fermi-Dirac distribution of electrodes at equilibrium, with $T$ being temperature and $k_\text{B}$ the Boltzmann constant.

As a result, the real part of the conductance,
$
	G_{\sigma\sigma^\prime}^{qq^\prime}(\omega)
	\equiv
	\textrm{Re}\acG_{\sigma\sigma^\prime}^{qq^\prime}(\omega)
	=
	-\textrm{Im}\GF{\opI_\sigma^q}{\opI_{\sigma^\prime}^{q^\prime}}_\omega^\textrm{r}/\omega
$,
can be expressed in units of the conductance quantum $G_0$ as
\begin{align}\label{eq:G_spin_def}
	\frac{G_{\sigma\sigma^\prime}^{qq^\prime}(\omega)}{G_0}
	=
	-
	\delta_{\sigma\sigma^\prime}
	\eta_q\eta_{q^\prime}
	2&\pi\sfA_0
	\cdot
	\frac{\Gamma_\sigma^L\Gamma_\sigma^R}{\Gamma_\sigma^\text{eff}}
	\cdot
	g_\sigma^o(\omega)
	\nonumber\\
	+\ &
	\pi\sfA_0
	\cdot
	\frac{\Gamma_\sigma^q\Gamma_{\sigma^\prime}^{q^\prime}}{\sqrt{\Gamma_\sigma^\text{eff}\Gamma_{\sigma^\prime}^\text{eff}}}
	\cdot
	g_{\sigma\sigma^\prime}^e(\omega)
	,
\end{align}
where $g_{\sigma}^o(\omega)$ and $g_{\sigma\sigma^\prime}^e(\omega)$ stand for dimensionless conductance contributions from the odd and even channels, respectively. These two functions can be then written as~\cite{Toth_Phys.Rev.B76/2007,Moca_Phys.Rev.B81/2010}
\begin{equation}\label{eq:g_odd}
	g_{\sigma}^o(\omega)
	=
	\frac{1}{2\omega}
	\int\!\intd\omega^\prime
	\frac{\sfA_\sigma^c(\omega^\prime)}{\sfA_0}
	\big[
	f(\omega^\prime-\omega)-f(\omega^\prime+\omega)
	\big],
\end{equation}
and
\begin{equation}\label{eq:g_even}
	g_{\sigma\sigma^\prime}^e(\omega)
	=
	-\frac{1}{\omega}
	\cdot
	\frac{\sfA_{\sigma\sigma^\prime}^{\mathcal{I}}(\omega)}{\hbar\rho\sfA_0}
	,
\end{equation}
with
$
	\sfA_\sigma^c(\omega)
	\equiv
	-\textrm{Im}\GF{c_\sigma^{}}{c_\sigma^\dagger}_\omega^\textrm{r}/\pi
$
denoting the spectral function of the OL,
$
	A_{\sigma\sigma^\prime}^{\mathcal{I}}(\omega)
	\equiv
	-\textrm{Im}\GF{\opII_\sigma^e}{\opII_{\sigma^\prime}^{e\dagger}}_\omega^\textrm{r}/\pi
$
being the spectral function for the dimensionless even current operator, and $\sfA_0=1/(\pi\Gamma)$ standing for the spectral function of a bare OL (that is, for $J=0$ which essentially corresponds to a single-level QD) at $\omega=0$ and for nonmagnetic electrodes.

In the following discussion we focus on the analysis of the left-right component of the dynamical conductance,
\begin{align}\label{eq:G_LR_def}
	G_{\text{P/AP}}(\omega)
	=\ &
	\sum_{\sigma\sigma^\prime}
	\big[
	G_{\sigma\sigma^\prime}^{LR}(\omega)
	\big]^{\text{P/AP}}
	\nonumber\\
	\equiv\ &
	G_{\text{P/AP}}^o(\omega)
	+
	G_{\text{P/AP}}^e(\omega)
	,
\end{align}
where the second line consists of two components of conductance, $G_c^o(\omega)$ and $G_c^e(\omega)$, related to the odd  and even transport channel, respectively.
For the parallel ($c=\text{P}$) and antiparallel ($c=\text{AP}$) magnetic configuration of the junction these quantities are defined as
\begin{equation}\label{eq:Gc_odd}
	G_c^o(\omega)
	=
	\frac{1}{2}
	\sum_\sigma
	\Lambda_\sigma^c
	\big[g_{\sigma}^o(\omega)\big]^{\!c}
	,
\end{equation}
and
\begin{equation}\label{eq:Gc_even}
	G_c^e(\omega)
	=
	\frac{1}{4}
	\sum_{\sigma\sigma^\prime}
	\Lambda_\sigma^c
	\Upsilon_{\sigma\sigma^\prime}^c
	\big[g_{\sigma\sigma^\prime}^e(\omega)\big]^{\!c}
	.
\end{equation}
When deriving the equation above, it was assumed as previously that both electrodes and tunnel barriers are identical, so that the factors $\Lambda_\sigma^c$ and $\Upsilon_{\sigma\sigma^\prime}^c$ determined by the magnetic configuration $c$ take the following form
\cite{Weymann_J.Appl.Phys.109/2011}
\begin{equation}
	\Lambda_\sigma^{\textrm{P}}
	=
	1+ \eta_\sigma p
	\quad
	\textrm{and}
	\quad
	\Lambda_\sigma^{\textrm{AP}}
	=
	1- p^2
	,
\end{equation}
with $\eta_{\uparrow(\downarrow)}=\pm 1$,
\begin{equation}
	\Upsilon_{\sigma\sigma^\prime}^\textrm{P}
	=
	\sqrt{\frac{1+\eta_{\sigma^\prime}p}{1+\eta_\sigma p}}
	\quad
	\textrm{and}
	\quad
	\Upsilon_{\sigma\sigma^\prime}^\textrm{AP}
	=
	\frac{1+\eta_\sigma p}{1+\eta_{\sigma^\prime} p}
	.
\end{equation}
%

\subsection{\label{sec:NRG}Numerical derivation of spectral functions}

As explained in the previous section, the spin-resolved dynamic conductance~(\ref{eq:G_spin_def}) is essentially determined by two types of spectral functions: one describing the OL, $\sfA_\sigma^c(\omega)$ [see Eq.~(\ref{eq:g_odd})], and the other for the dimensionless even current operator, $A_{\sigma\sigma^\prime}^{\mathcal{I}}(\omega)$ [see Eq.~(\ref{eq:g_even})]. In general, derivation 
of these two quantities in the Kondo regime ---corresponding to a strong tunnel coupling between a spin impurity and conducting electrons--- is a non-trivial task. In the present work we use for this purpose the Wilson's numerical renormalization group method~\cite{Wilson_RMP.47/773,Bulla_RMP.80/395,Legeza_DMNRGmanual,Toth_Phys.Rev.B78/2008}.
The main idea of the NRG technique relies on
the logarithmic discretization of the conduction band with a discretization parameter $\Lambda$, so that transport properties of a system can be resolved on energy scales logarithmically approaching the Fermi level. Next, such a discretized model is mapped onto a semi-infinite chain with exponentially decaying hoppings and the first site being coupled to a spin impurity.

Relating this idea to our model of a magnetic molecule,
the full NRG Hamiltonian of the system under consideration takes the following form:
\begin{equation}
	\Ham_\text{NRG} 
	= 
	\Ham_\text{MQD} + \Ham_\text{chain} + \Ham_\text{MQD-chain}
\end{equation}
where the second term, representing the Wilson chain, is given by
\begin{equation}
	\Ham_\textrm{chain} 
	=
	\sum_{n=0}^\infty
	\sum_\sigma
	V_{n} 
	\big[ \opf^{\dag}_{n \sigma} \opf_{n+1 \sigma}^{} 
	+ 
	\opf_{n+1 \sigma}^{\dag} \opf_{n \sigma}^{} \big]
	.
\end{equation}
Here, the operators $f^{\dag}_{n \sigma}$ and $f_{n \sigma}$ describe the $n$th site of the Wilson chain, while electron hopping between neighboring sites of this chain is characterized by matrix elements $V_n$.
Furthermore, the coupling between the MQD and the first site of the semi-infinite chain is formulated by means of the effective hybridization function~$\Gamma_\sigma^\text{eff}$, cf.~Eq.~(\ref{eq:H_tun_eff}), as
\begin{equation}
	\Ham_\textrm{MQD-chain} 
	=
	\sum_{ \sigma} 
	\sqrt{\frac{\Gamma_{\sigma}^\text{eff}}{\pi\rho}} 
	\big[ \opc^{\dag}_{\sigma} \opf_{0 \sigma}^{} 
	+ 
	\opf_{0 \sigma}^{\dag} \opc_{\sigma}^{} \big]
	.
\end{equation}
Having defined the NRG Hamiltonian we continue with the next step of this method
which is the iterative diagonalization of the chain.
In calculations we took the discretization parameter $\Lambda = 2$
and $2048$ states were kept after each step of the iteration.
Moreover, to improve the accuracy of calculations,
we averaged the spectral data over $N_z = 8$ different discretization meshes~\cite{OliveiraPhysRevB.49.11986}.

\section{\label{sec:QD}The case of a quantum dot ($J=0$)}

To set the ground for a discussion of frequency-dependent transport through a magnetic molecule in the Kondo regime, we first introduce some general concepts by considering 
the instructive example of a single-level quantum dot (QD). In particular, this limiting case is obtained by assuming $J=0$, as mentioned in Sec.~\ref{sec:Model}.
Importantly, the essential concepts that we learn from the analysis of the transport through a QD
will prove very useful for understanding of transport mechanisms
relevant in the case of a more complex system such as a MQD.
We note that whereas stationary transport through QDs strongly coupled to ferromagnetic contacts
have already been the subject of extensive experimental~\cite{Pasupathy_Sc.306/86,
Heersche_PRL.96/017205,Hamaya_PRB.77/081302,Hauptmann_PRB.4/373-376,Gaass_Phys.Rev.Lett.107/2011}
and theoretical~\cite{Martinek_Phys.Rev.Lett.91/2003_127203,Martinek_Phys.Rev.Lett.91/2003_247202,
Swirkowicz_Phys.Rev.B73/2006,Weymann_Phys.Rev.B83/2011} studies,
the problem of time-dependent transport has attracted less attention so far~\cite{Weymann_J.Appl.Phys.109/2011}.

\subsection{\label{sec:Parameters}Parameter space}

To begin with, let us recall the key conditions which have to be met in order to observe a formation of the Kondo resonance.
First of all, the dot has to be occupied by a single electron so that spin-flip processes responsible for the increase of conductance can take place. This is ensured by adjusting a gate voltage to keep the position of the dot level $\varepsilon$ in the range: $-1\lesssim\varepsilon/U\lesssim0$.
Secondly, the temperature~$T$ of the system should be lower than
the Kondo temperature~$\TK$~\cite{Hewson_book},
which represents the characteristic energy scale of the effect under consideration.
In our system the Kondo temperature is estimated from the temperature dependence of zero-frequency conductance~$G(\omega=0,T)$ as the half-width at the half-maximum of the normalized linear conductance~$G(\omega=0,T)/G(\omega=0,T=0)$ at the particle-hole symmetry point ($\varepsilon/U=-1/2$), so that we find $\TK/W=0.002$, with temperature given in energy units ($k_\text{B}\equiv1$) and the conduction band half-width $W\equiv1$ taken as an energy unit.
This value of the Kondo temperature, henceforth referred to as $\TKqd\equiv\TK(J=0)$, is obtained by assuming the following parameters of the system in our calculations: the Coulomb energy associated with the double occupation of the dot is $U/W=0.4$ and the tunnel coupling between  electrodes and a QD is $\Gamma/U=0.1$. From now on, whenever useful, $\TKqd$ will be employed as a reference energy scale.
Furthermore, if not stated otherwise, the spin polarization coefficient of electrodes is  $p=0.5$.
Numerical results are presented here for $T=0$ and  a wide range of frequencies. As will be discussed below, the most subtle effects are observed in the regime of low frequencies.
On the other hand, in the limit of high frequencies the Coulomb interaction starts to play a dominant role and the Kondo effect becomes suppressed.

\subsection{Numerical results and discussion}

%
\begin{figure}[t]
	\includegraphics[width=0.99\columnwidth]{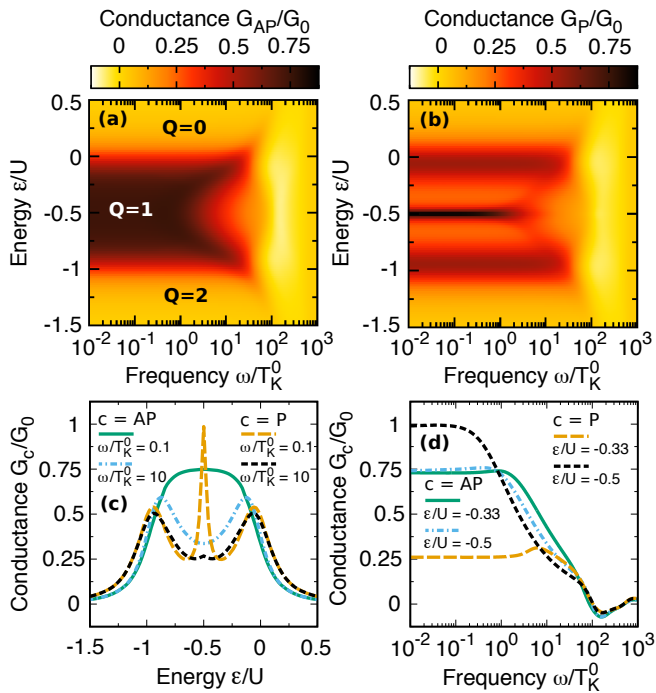}
	\caption{(color online)
	Dynamical conductance $G_c(\omega)$ of a quantum dot (QD), i.e., for $J=0$, shown as a function of frequency $\omega$ and the position of the QD level $\varepsilon$ for the antiparallel ($c=\text{AP}$) and parallel ($c=\text{P}$) magnetic configuration of electrodes. Recall that the normalization factor $G_0\equiv2e^2/h$ denotes the conductance quantum. 
	\emph{Bottom panels} present relevant cross-sections of (a) and~(b): for indicated values of~$\omega$ in (c) and for chosen values of~$\varepsilon$ in (d).  
	Note that the scales of color bars in (a)-(b) are purposefully chosen the same to facilitate the comparison of the results. In fact, in~(b) the maximum value of conductance is $G_\text{P}(\omega)/G_0=1$, as can be seen in (c)-(d).
	The parameters are: $U=0.4$ (in units of band half-width),
	$\Gamma / U = 0.1$, $\TKqd/U = 0.05$ and $p=0.5$
	($\hbar=k_{\rm B}\equiv 1$).
	\label{fig2}
	}
\end{figure}

Figure~\ref{fig2} presents the dependence of dynamical conductance $G(\omega)$, see Eq.~(\ref{eq:G_LR_def}), on frequency $\omega$ and the position of a QD level $\varepsilon$ for the parallel (a) and antiparallel (b) orientation of spin moments of electrodes. In each case, one can observe the onset of the Kondo effect at low frequencies $\omega<\TKqd$ for $\varepsilon$ corresponding to a single occupation ($Q=1$) of the QD, though the behavior of conductance is significantly different depending on the magnetic configuration of electrodes.
Specifically, when $\omega\ll \TKqd$,
in the antiparallel configuration $G_\text{AP}(\omega)/G_0\rightarrow 1-p^2=0.75$ for $-1\lesssim\varepsilon/U\lesssim0$, as indicated by the solid line in Fig.~\ref{fig2}(c), whereas in the parallel configuration the conductance reaches the unitary limit, $G_\text{AP}(\omega)/G_0\rightarrow 1$, only at the particle-hole symmetry point ($\varepsilon/U=-1/2$), see the long-dashed line in Fig.~\ref{fig2}(c).
This striking qualitative difference between the dependence of $G(\omega)$ on the position of the QD level~$\varepsilon$ for the antiparallel and parallel magnetic configuration stems from the presence of an effective exchange field in the latter case~\cite{Martinek_Phys.Rev.Lett.91/2003_127203,Sindel_Phys.Rev.B76/2007}. This field occurs as a result of a spin asymmetry of the effective tunnel coupling $\Gamma_{\uparrow(\downarrow)}^\text{eff}\propto 1\pm p$ and it leads to the spin splitting of the~QD level $\delta\varepsilon_\text{ex}\propto \ln|\varepsilon/(\varepsilon+U)|$.
On the other hand, if the dot is occupied by an even number of electrons, that is, $Q=0$ for $\varepsilon/U\gtrsim0$ and $Q=2$ for $\varepsilon/U\lesssim-1$, conductance becomes suppressed since the spin exchanging cotunneling processes play no role.
These results are in agreement with previous studies for stationary ($\omega=0$)
spin-dependent transport through a single-level QD~\cite{Weymann_Phys.Rev.B83/2011}.

In the opposite limit of large frequencies $\omega\gg\TKqd$, one can notice that the Kondo effect gets attenuated as soon as $\omega\gtrsim\TKqd$ regardless of the magnetic configuration of electrodes. 
To understand this behavior, one should note that the presence of periodically varying in time bias voltage essentially corresponds to pumping energy into the system, allowing thus for its excitation.
Consequently, when $\omega$ significantly exceeds~$\TK$ and starts approaching the limit of $\omega\sim U$ the qualitative difference between transport features for the two magnetic configurations becomes negligible, see Fig.~\ref{fig2}(d), as the occupation of the dot by two electrons is energetically accessible. In such a regime conductance can even become negative,
which can be attributed to the dominant role of the Coulomb interaction
and large charge fluctuations present for~$\omega \approx U$~\cite{Weymann_J.Appl.Phys.109/2011}.

We use the above model of a single-level QD as a starting point for discussion of a more complex system, that is, a large-spin magnetic molecule, which is the subject of the present paper. Knowing the generic frequency-dependent transport characteristics of QDs in the Kondo regime, we will be able thus to discern these from other subtle and non-trivial effects occurring only for large-spin systems, which we consider in the next section.

\section{\label{sec:Large_spin}The case of a large-spin\\ magnetic molecule~($J\neq0$)}

In the present section,  we extend our analysis of frequency-dependent transport in the Kondo regime to large-spin systems ---specifically, to magnetic molecules modeled as described in Sec.~\ref{sec:Model}. For this purpose, unlike in the previous section, we assume here the finite value of the exchange interaction parameter $J\neq0$, so that magnetic molecules of interest are represented by Hamiltonian~(\ref{eq:H_MQD}). To make the following discussion comprehensible, the value of an internal spin  (magnetic core) is taken to be $S=2$, which allows for a comparison of the present results with the case of stationary transport studied in such a system before~\cite{Misiorny_Phys.Rev.Lett.106/2011,Misiorny_Phys.Rev.B84/2011,Misiorny_Phys.Rev.B86/2012,Misiorny_Phys.Rev.B90/2014}. We note that currently the main aim of the work is to explain the subtle effects, occurring  due to the generic properties of a magnetic molecule, such as a large spin and magnetic anisotropy, which are visible in the frequency-dependent transport characteristics.

However, before we get down to the effect of magnetic anisotropy, at first we study the frequency-dependent transport through a \emph{spin-isotropic} molecule (i.e., for $D=0$). In our calculations we assume the magnitude of the exchange coupling between the OL and the internal spin to be $|J|/\TKqd=2.25$. Moreover, no constraint regarding the sign of $J$ is  imposed, so that two types of the interaction are generally considered: ferromagnetic $(\text{FM},\, J>0)$ or antiferromagnetic $(\text{AFM},\, J<0)$.
Since the value of the parameter~$J$ in relation to the Kondo temperature~$\TKqd$ plays a decisive role in formation of the the Kondo effect~\cite{Misiorny_Phys.Rev.B86/2012}, below we motivate the relevance of this specifically chosen value of $J$. To begin with, recall that regardless of the type of the $J$-coupling, the well-pronounced Kondo resonance is always observed as long as $|J|\ll \TKqd$. In this regime, the spin $\oper{\bm{s}}$ of an electron in the OL of the molecule becomes screened by conduction electrons from electrodes, and the presence of the internal spin essentially does not qualitatively affect the Kondo resonance. 
On the other hand, if the condition  $|J|\gtrsim\TKqd$ is met, as temperature of the system is decreased for $J>0$ one observes the \emph{underscreened Kondo effect}
\cite{Koller_Phys.Rev.B72/2005,Roch_Phys.Rev.Lett.103/2009,Weymann_Phys.Rev.B81/2010,Cornaglia_Europhys.Lett.93/2011,
Misiorny_Phys.Rev.B86/2012_UK},
while for $J<0$ the so-called \emph{two-stage Kondo effect} takes place
\cite{Wiel_Phys.Rev.Lett.88/2002,Posazhennikova_Phys.Rev.B75/2007,
Zitko_J.Phys.:Condens.Matter22/2010,Misiorny_Phys.Rev.B86/2012,Wojcik_PhysRevB.91.134422/2014}.
In the latter situation, with diminishing temperature, first, the OL spin is screened and the conductance increases, while for even lower temperature this new effective Fermi sea screens also the internal spin (which is ensured by the AFM $J$-coupling) leading to the suppression of conductance. Consequently, assuming the parameter $J$ to be slightly larger than the Kondo temperature, one predicts that at $T=0$ the Kondo effect should occur only for the FM $J$-coupling, whereas for the AFM case conductance of the system should be suppressed.


%
\begin{figure}[t]
	\includegraphics[width=1\columnwidth]{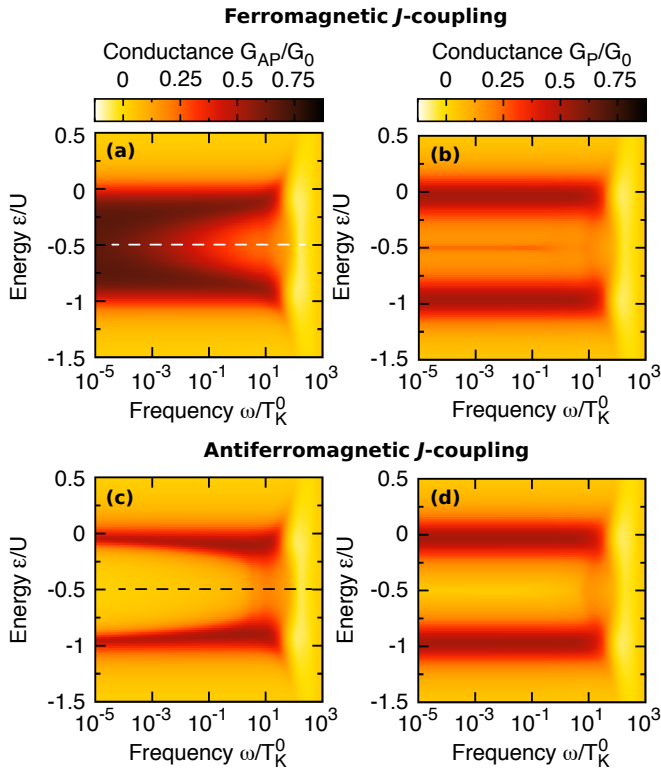}
	\caption{
	(color online)
	Frequency-dependence conductance of a \emph{spin-isotropic} ($D=0$) magnetic
	molecule plotted as a function of the OL position~$\varepsilon$
	for the antiparallel (AP, left column) and parallel (P, right column) magnetic configuration of electrodes.
	The \emph{top} [\emph{bottom}] \emph{panel} corresponds to the FM ($J>0$) [AFM ($J<0$)] exchange coupling between the OL spin and the internal one.
	When comparing with Figs.~\ref{fig2}(a)-(b), note that here the frequency range extends towards lower frequencies.
	Moreover, recall that the Kondo temperature $\TKqd$ used here as an energy scaling factor is defined for a single-level QD.
      Except $|J|/\TKqd=2.25$, other parameters of the system are the same as in Fig.~\ref{fig2}.
	\label{fig3}
	}
\end{figure}

The frequency-dependent conductance of a spin-isotropic molecule
plotted as a function of $\varepsilon$ and $\omega$
for two different types of the $J$-coupling
is shown in~Fig.~\ref{fig3}.
One can immediately compare the effect of a large internal spin and its coupling to the
OL spin on frequency-dependent transport
with the case of a single-level QD in Sec.~\ref{sec:QD}, see Fig.~\ref{fig2}. 
It can be seen that for the~FM~\mbox{$J$-coupling} the behavior of the conductance in the low-frequency limit ($\omega/\TKqd\rightarrow0$) agrees qualitatively with that of a QD, cf. Figs.~\ref{fig3}(a)-(b) with Figs.~\ref{fig2}(a)-(b). Importantly, while in the parallel magnetic configuration
[Fig.~\ref{fig3}(b)] this behavior is preserved until $\omega\approx\TKqd$, in the antiparallel configuration a suppression of conductance  occurs for $-1\lesssim\varepsilon/U\lesssim0$ long before such a limit of frequency is reached [Fig.~\ref{fig3}(a)]. To understand this difference, first, we recall that the Kondo temperature $\TKqd$, to which frequency is scaled, refers in fact to a single-level QD. Since one generally expects that suppression of conductance occurs for $\omega\gtrsim\TK$, one can conclude that in the case under discussion the Kondo temperature $\TK$ has been reduced, as compared to the case of a QD~\cite{Misiorny_Phys.Rev.Lett.106/2011}.
Thus, this effect can be basically attributed to the presence of the exchange coupling of the OL to an internal spin.
In fact, the suppressed Kondo temperature can be also observed 
in the case of parallel configuration, which can be directly inferred
from the width of the Kondo peak at $\varepsilon=-U/2$ as a function of $\varepsilon$,
which is now much smaller compared to the QD case,
cf. Figs.~\ref{fig3}(b) with Figs.~\ref{fig2}(b).
This implies that a smaller detuning from particle-hole symmetry point,
that is, a smaller exchange-field-induced splitting of the OL,
is necessary to destroy the Kondo resonance.
In addition, one can also note that the height of the Kondo peak
in the parallel configuration is much lower compared the case of $J=0$.
This feature results directly from the presence of quadrupolar
field, as will be discussed in greater detail in the next sections.

On the contrary, a completely different behavior of conductance
is observed for the AFM $J$-coupling, where, owing to the two-stage Kondo effect,
a significant suppression of conductance takes place
in both magnetic configurations, see 
Figs.~\ref{fig3}(c) and (d).
Note, however, that this suppression is more effective 
in the case of antiparallel configuration,
since in the parallel configuration the presence of exchange fields
obscures the second stage of Kondo screening.

\subsection{\label{sec:Mag_aniso}The effect of magnetic anisotropy}

\begin{figure}[t]
	\includegraphics[width=0.99\columnwidth]{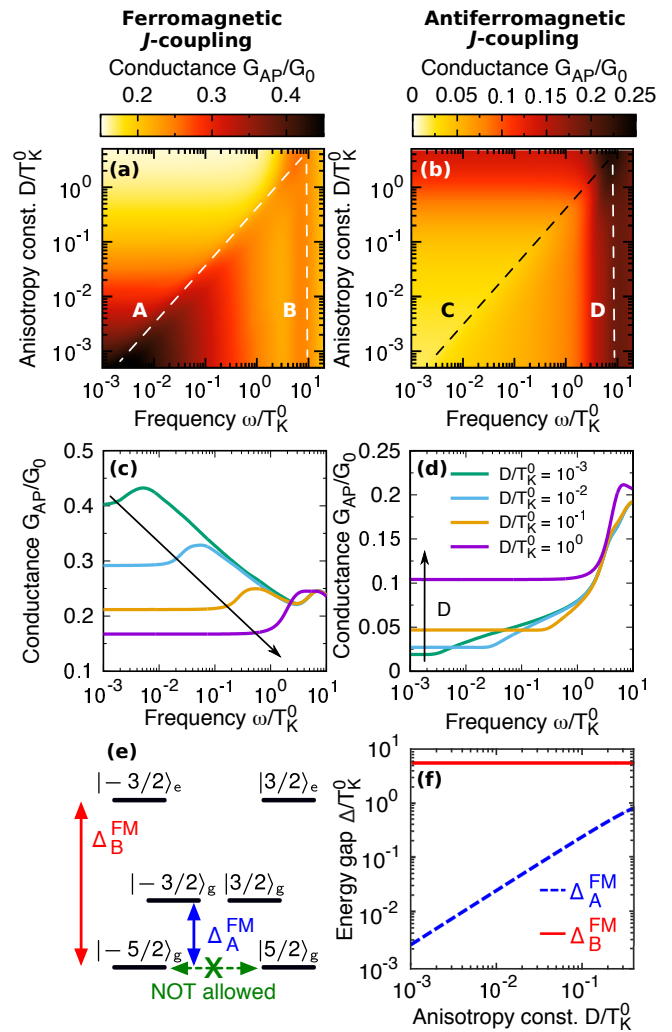}
	\caption{
	(color online)
	The effect of uniaxial magnetic anisotropy on frequency-dependent conductance of a large-spin molecule.
	\emph{Top panels} [(a)-(b)]: Conductance density maps for the antiparallel magnetic configuration of electrodes shown at the particle-hole symmetry point ($\varepsilon=-U/2$) as functions of frequency $\omega$ and the uniaxial magnetic anisotropy constant $D$ in the case of the ferromagnetic (a) and antiferromagnetic (b) $J$-coupling.
	\emph{Middle panels} [(c)-(d)]: Cross-sections of the respective panels above for selected values of~$D$ as specified in (d). Black arrows indicate here the direction of increasing~$D$.
	(e) Schematic representation of a few low-energy states of a free-standing molecule participating in spin-exchange electron tunneling processes \via the OL leading to the Kondo effect. For further details see Sec.~\ref{sec:Mag_aniso}.
	(f) Dependence of the excitation gaps marked in (e) on the uniaxial magnetic anisotropy~$D$.
	All other parameters are the same as in Fig.~\ref{fig3}.
	\label{fig4}
	}
\end{figure}

The physical picture of frequency-dependent transport established above for a spin-isotropic molecule becomes more complex if the internal spin exhibits additionally spatial preference for its orientation ---namely, we include now the uniaxial magnetic anisotropy ($D\neq0$). To keep the discussion focused on key aspects of the problem, let us for the moment concentrate only on the case of the antiparallel magnetic configuration of electrodes and the position of the OL corresponding to the particle-hole symmetry point ($\varepsilon=-U/2$). For a spin-isotropic molecule this situation corresponds to cuts along the thin dashed lines in Figs.~\ref{fig3}(a,c), and in Figs.~\ref{fig4}(a)-(b) we present how these evolve with increasing $D$. The motivation for choosing the antiparallel configuration stems from the fact that in such a case no effective spintronic exchange fields arise~\cite{Misiorny_NaturePhys.9/2013}, which, in turn, allows for studying  effects exclusively originating from the generic properties of the molecule interconnecting electrodes. The effect of such fields will be discussed in detail in the next section (Sec.~\ref{sec:Ex_fields}).

Analyzing the evolution of frequency-dependent conductance $G_\text{AP}(\omega)$ in response to larger and larger magnetic anisotropy $D$, as shown in Fig.~\ref{fig4}, one can see a markedly different behavior of the conductance for the FM- (a,c) and AFM-type (b,d) of the $J$-coupling. 
In particular, this occurs for low frequencies ($\omega\ll\TKqd$) where for the FM $J$-coupling $G_\text{AP}(\omega)$ becomes suppressed for large values of  $D$ and a distinct resonance emerges, delineated by the dashed thin line A in Fig.~\ref{fig4}(a) [see also Fig.~\ref{fig4}(c)]. 
On the other hand, for the AFM $J$-coupling under the same conditions $G_\text{AP}(\omega)$ is enhanced and a kink forms, highlighted in Fig.~\ref{fig4}(b) by the dashed thin line C  [see also Fig.~\ref{fig4}(d)].
Importantly, the position of this peak/kink shifts towards higher frequencies as the magnetic anisotropy becomes stronger. This indicates that the spin-exchange processes underlying the Kondo effect occur between some ground and excited states with the energy gap between them determined by the magnetic anisotropy parameter~$D$. To elucidate more this statement,
let us focus on the case illustrated in Figs.~\ref{fig4}(a,c), that is, for the FM $J$-coupling.

As one can notice, there are two characteristic resonances indicated by thin dashed lines A and B. As already mentioned above, resonance A starts at low frequencies~$\omega$ for vanishingly small magnetic anisotropy $D$ and its position with respect to $\omega$ depends linearly on~$D$ (mind logarithmic scales both for $\omega$ and $D$). On the other hand, resonance B develops at  $\omega/\TKqd \approx 10^1$.
The physical mechanism leading to formation of these two resonances can be qualitatively understood by considering the lowest energy spectrum of a  free-standing magnetic molecule, that is, we neglect the renormalization effects due to strong tunnel-coupling of the molecule to electrodes. 
First of all, it is worth noticing that the coupling between
the OL spin and the internal spin results in
decomposition of the MQD states (with the OL occupied by a single electron)
into two spin multiplets $S\pm1/2$.
Then, the sign of the parameter $J$ representing this coupling,
see Eq.~(\ref{eq:H_J}), determines which of these two multiplets
is characterized by lower energy ---in the situation under discussion
for $J>0$ (FM) it is $S+1/2$.
In the absence of magnetic anisotropy ($D=0$)
all the states within each spin multiplet are degenerate,
so that as long as temperature is low enough one observes
the Kondo effect at low frequencies, see Fig.~\ref{fig3}(a).
Next, adding the uniaxial magnetic anisotropy to the problem
results in the additional splitting of the states within
each spin multiplet and also modifies the excitation gaps
between states belonging to different multiplets~\cite{Misiorny_Phys.Rev.B84/2011}.

In the case under consideration (i.e., $J>0$),
at $\varepsilon=-U/2$, $D\neq0$ and $T=0$,
the states of a singly occupied MQD that
primarily contribute to a formation of the Kondo effect are:\\
(i)~from the spin multiplet $S+1/2$ being \emph{lower} in
energy (referred to as the `\emph{ground}' multiplet, and marked by the subscript `g')
\begin{equation}\label{eq:States_L}
	\left\{
	\begin{aligned}
	\Ket{S_z^{\text{tot}}=\pm5/2}_\text{g}
	=\  &
	\Ket{\pm1/2}_{\textrm{OL}} 
	\otimes 
	\Ket{\pm 2}_{\textrm{S}},
	\\[5pt]
	\Ket{S_z^{\text{tot}}=\pm3/2}_\text{g}
	=\  &
	\mathcal{X}^{\text{g}}_{\pm}\,
	\Ket{\pm1/2}_{\textrm{OL}}
	\otimes 
	\Ket{\pm1}_{\textrm{S}}
	\\
	+\ &
	\mathcal{Y}^{\text{g}}_{\pm}\, 
	\Ket{\mp1/2}_{\textrm{OL}}
	\otimes 
	\Ket{\pm2}_{\textrm{S}}
	,
	\end{aligned}
	\right.
\end{equation}
(ii)~from the spin multiplet $S-1/2$ being \emph{higher} in energy (referred to as the `\emph{excited}' multiplet, and marked by the subscript `e')
\begin{align}\label{eq:States_H}
	\Ket{S_z^{\text{tot}}=\pm3/2}_\text{e}
	=\  &
	\mathcal{X}^{\text{e}}_{\pm} \, 
	\Ket{\pm1/2}_{\textrm{OL}}
	\otimes 
	\Ket{\pm1}_{\textrm{S}}
	\nonumber\\
	+\ &
	\mathcal{Y}^{\text{e}}_{\pm}\, 
	\Ket{\mp1/2}_{\textrm{OL}}
	\otimes 
	\Ket{\pm2}_{\textrm{S}}
	.
\end{align}
Here, $\Ket{\!\bullet\!}_{\text{OL(S)}}$ stands for the spin state of OL (magnetic core), whereas 
$\mathcal{X}^{\text{g(e)}}_{\pm}$ and $\mathcal{Y}^{\text{g(e)}}_{\pm}$ are effective Clebsch-Gordon coefficients, which non-trivially depend on~$J$ and~$D$~\cite{Misiorny_Phys.Stat.Sol.B246/2009}. For the sake of notational clarity, below we use $\Ket{S_z^{\text{tot}}=m}_\text{g(e)}\equiv\Ket{m}_\text{g(e)}$.
We recall that the Kondo effect arises due to conduction-electron co-tunneling processes which lead to spin exchange (flip) in the OL.  As a result, in the considered system such spin-exchange processes correspond to the ground-to-excited-state transitions within the ground multiplet, $\Ket{5/2}_{\text{g}}\leftrightarrow\Ket{3/2}_{\text{g}}$  and $\Ket{\!-5/2}_{\text{g}}\leftrightarrow\Ket{\!-3/2}_{\text{g}}$, as well as between the ground and excited multiplets, $\Ket{5/2}_{\text{g}}\leftrightarrow\Ket{3/2}_{\text{e}}$ and $\Ket{\!-5/2}_{\text{g}}\leftrightarrow\Ket{\!-3/2}_{\text{e}}$, with the relevant energy gaps  $\Delta_\text{A}^\text{FM}$ (for $\text{g}\leftrightarrow\text{g}$) and $\Delta_\text{B}^\text{FM}$ (for $\text{g}\leftrightarrow\text{e}$) as indicated in Fig.~\ref{fig4}(e). Note, however, that the direct ground-to-ground-state transition $\Ket{\!-5/2}_{\text{g}}\leftrightarrow\Ket{5/2}_{\text{g}}$ is prohibited as a single conduction electron is not able to change the state of the magnetic core by $2S$, see Eq.~(\ref{eq:States_L}). 
In the regime of $D\ll J$, the two energy gaps can be estimated to be~\cite{Misiorny_Phys.Rev.B84/2011}:
\begin{equation}\label{eq:Delta_A}
	\Delta_{\textrm{A}}^{\textrm{FM}} 
	\approx 
	2SD 
	\bigg[1 - \frac{2(J-D)}{(2S+1)(J-2D)} \bigg]
	,
\end{equation}
and
\begin{equation}\label{eq:Delta_B}
	\Delta_{\textrm{B}}^{\textrm{FM}} 
	\approx
	\frac{2S+1}{2}
	J
	.
\end{equation}
One can easily see that
$\Delta_{\textrm{A}}^{\textrm{FM}}\propto D$
and $\Delta_{\textrm{B}}^{\textrm{FM}}\propto J$, see Fig.~\ref{fig4}(f).
Consequently, whereas in the limit of stationary transport ($\omega\rightarrow0$) the Kondo effect can take place only if $\Delta_{\textrm{A}}^{\textrm{F}}\lesssim\TK$~\cite{Misiorny_Phys.Rev.Lett.106/2011,Misiorny_Phys.Rev.B84/2011}, as otherwise the spin exchange processes are energetically not allowed, for $\omega\neq0$ such transitions become possible if the resonant condition $\omega\approx\Delta_{\textrm{A}}^{\textrm{FM}}$ or $\omega\approx\Delta_{\textrm{B}}^{\textrm{FM}}$ is met. It essentially means that the energy pumped to the MQD matches the excitation energy between the molecular states participating in the process of spin exchange. This, in turn, corresponds to resonances A and B in Fig.~\ref{fig4}(a). The former one is associated then with the transition between states within the ground spin multiplet, and thus, it depends linearly~on~$D$, see Eq.~(\ref{eq:Delta_A}). The latter resonance, on the other hand, originates from transition between states belonging to the ground and excited spin multiplets, so that its position is determined solely by the magnitude of the $J$-coupling, but not by $D$, see Eq.~(\ref{eq:Delta_B}). 
A small quantitative mismatch between energy gaps plotted in Fig.~\ref{fig4}(f) and the position of resonances A and B in Fig.~\ref{fig4}(a) stems from the fact that above we considered the states of a free-standing molecule, neglecting thus the renormalization of their energies, which is systematically included in the NRG calculations.
The occurrence of the characteristic features in Fig.~\ref{fig4}(b), that is the kink~/~peak marked by~C~/~D, can be explained using the analogous line of argumentation.

Finally, note that the upper limit of frequencies plotted in Fig.~\ref{fig4}(a)-(b) is much smaller than $U$. For even larger~$\omega$ the same behavior as in Figs.~\ref{fig3}(a,c), determined by the Coulomb interaction, is observed.

\subsection{\label{sec:Ex_fields}The effect of the exchange field}

In this section we focus on another peculiar aspect of spin-polarized frequency-dependent transport, namely, on the effect of effective exchange fields~\cite{Martinek_Phys.Rev.Lett.91/2003_127203,Konig_Lect.NotesPhys.658/2005,Misiorny_NaturePhys.9/2013}. Importantly, these can  significantly affect transport properties of a nanosystem, e.g., manifesting by gate-voltage-dependent splitting of the Kondo resonance~\cite{Martinek_Phys.Rev.B72/2005} or even inducing spin anisotropy in a generically spin-isotropic system~\cite{Misiorny_NaturePhys.9/2013}. 
As highlighted at the beginning of the previous section, such fields do not occur in the antiparallel magnetic configuration of electrodes. This is not the case for the parallel magnetic configuration, and we expect that this should be visible especially in the behavior of features A and C in Figs.~\ref{fig4}(a)~and~(b), respectively. 
For the purpose of the following discussion, to make these two features more pronounced, we introduce here a new normalization of conductance $G(\omega)$ with respect to its value at $\omega=0$.

\begin{figure}[t]
	\includegraphics[width=0.99\columnwidth]{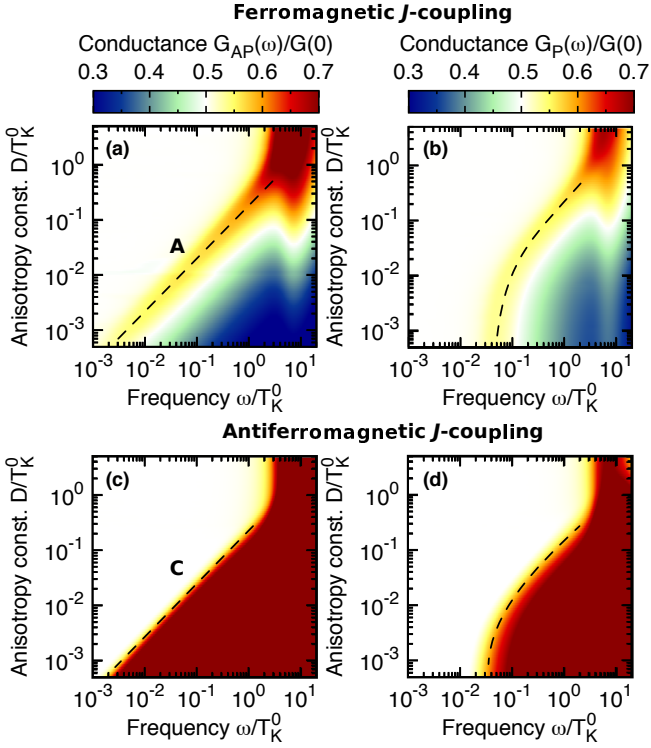}
	\caption{
	(color online)
	Dynamical conductance $G_c(\omega)$ scaled with respect to its value at $\omega=0$ plotted as a function of frequency $\omega$ and the uniaxial magnetic anisotropy constant~$D$ for $\varepsilon=-U/2$. The \emph{left} (\emph{right}) \emph{column} corresponds to the antiparallel (parallel) [$c=\text{AP\,(P)}$] magnetic configuration of electrodes.
	\emph{Top/Bottom panels} represent the case of the FM/AFM $J$-coupling.
	All remaining parameters of the system are the same as in Fig.~\ref{fig3}.  
	\label{fig5}
	}
\end{figure}
\begin{figure}[t]
	\includegraphics[width=0.99\columnwidth]{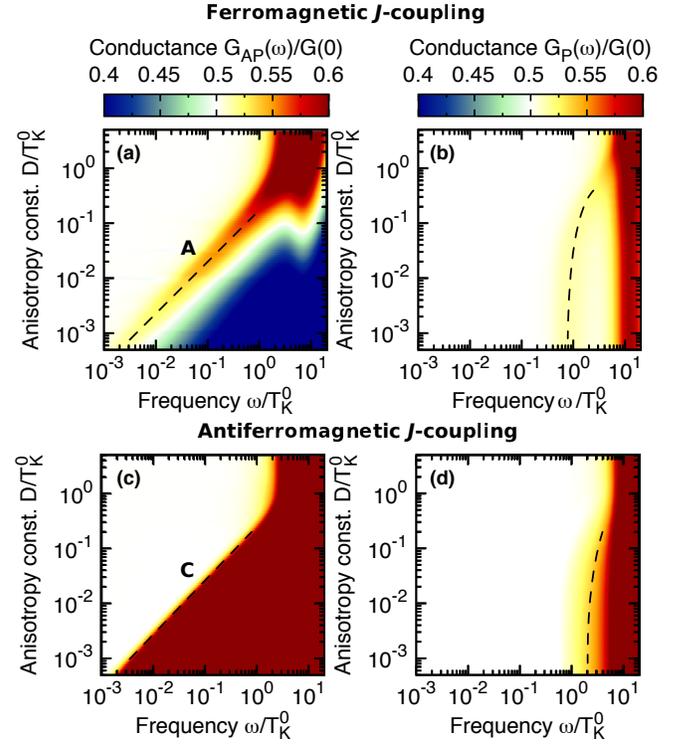}
	\caption{
	(color online)
    Analogous to Fig.~\ref{fig5} except that now the conductance is presented for $\varepsilon=-U/3$.
	\label{fig6}
	}
\end{figure}

To begin with, let us first replot Figs.~\ref{fig4}(a)-(b) using the new normalization scheme (the left column of Fig.~\ref{fig5}) and supplement these by the plots representing the case of the parallel magnetic configuration (the right column of Fig.~\ref{fig5}). It can be seen that the peak A and step C no longer scale linearly with magnetic anisotropy, as explained in the previous section, once the magnetic configuration of a device is switched to the parallel one. Whereas for large values of $D$ this linear dependence is still preserved, for smaller and smaller~$D$ these features saturate at specific values of $\omega$. In other words, even if the molecule is spin-isotropic ($D\rightarrow0$), there is always an excitation gap to induce the spin-exchange processes.
Since Fig.~\ref{fig5} is obtained for the particle-hole symmetry point ($\varepsilon=-U/2$), where only the quadrupolar exchange field is present~\cite{Misiorny_NaturePhys.9/2013}, the observed behavior of features A and C in Figs.~\ref{fig5}(b,d) we attribute to the effective spintronic magnetic anisotropy induced by such a field. 
One should understand this as follows: Eq.~(\ref{eq:Delta_A}) describing the energy gap $\Delta_{\textrm{A}}^{\textrm{FM}}$ can be generalized to include the spintronic component of magnetic anisotropy $\Ds$ (i.e., the effective quadrupolar exchange field) by making the substitution $D\mapsto D+\Ds$. Thus, $D$ can be now interpreted as the intrinsic component of magnetic anisotropy, that is, present also when a molecule is not attached to electrodes. It goes without saying that even for a spin-isotropic molecule in a magnetic junction it may become necessary to apply a bias voltage of the resonant frequency to observe the Kondo effect.

The voltage of even higher frequency is required if one moves away from the particle-hole symmetry point where the effective dipolar exchange field starts playing a dominating role.  Figure~\ref{fig6} illustrates how features~A and~C look at~$\varepsilon=-U/3$, that is, away from the symmetry point. One can see that whereas for the antiparallel magnetic configuration the qualitative change is negligible, for the parallel one the saturation of the two features at higher frequencies is more distinct. 
This shift of features~A and~C towards higher frequencies is associated with the increase of energy gaps between relevant molecular states participating in the spin-exchange processes due to a Zeeman component from the dipolar exchange field.
We remind that no external magnetic field is applied and
the dipolar field under consideration results solely from spin-dependent electron tunneling.
Such tunneling gives rise to the associated spin-resolved level renormalization
of the orbital level, which has different sign for each spin species
and effectively splits the level of the molecule.

\begin{figure}[t]
	\includegraphics[width=0.99\columnwidth]{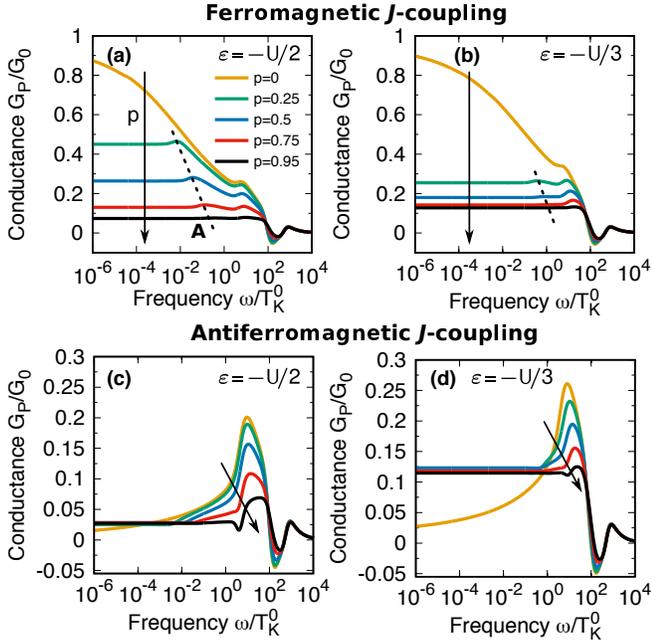}
	\caption{
	(color online)
	The effect of spin-polarization of electrodes $p$ on frequency-dependent conductance
	in the case of parallel magnetic configuration and a \emph{spin-isotropic} molecule ($D=0$).
	The \emph{left} (\emph{right}) \emph{column} represents the case of $\varepsilon=-U/2$ ($\varepsilon=-U/3$).
	Black arrows indicate the direction of increasing $p$, while finely dashed lines in (a)-(b) serve as a guide for eyes marking the position of the resonance A, cf. Fig.~\ref{fig4}(a).
	Other parameters of the system are the same as in Fig.~\ref{fig3}.
	\label{fig7}
	}
\end{figure}

Finally, to corroborate the above findings,
we focus on the behavior of the dynamical conductance
for different values of spin polarization $p$
in the case of a \emph{spin-isotropic} molecule
and parallel magnetic configuration.
This implies that no intrinsic component of magnetic anisotropy is present
($D=0$), but the effective exchange fields are still in action, see Fig.~\ref{fig7}.
The left column of Fig.~\ref{fig7} corresponds to the particle-hole symmetry point $\varepsilon=-U/2$,
where only the quadrupolar field can arise,
while in the right column we present results at $\varepsilon=-U/3$,
where both the dipolar and quadrupolar fields are present.

Let us first concentrate on the case of the FM $J$-coupling shown in Figs.~\ref{fig7}(a)-(b).
One can immediately notice that when the electrodes are nonmagnetic ($p=0$),
in the stationary limit  $\omega\rightarrow0$ conductance approaches the unitary
value of the conductance quantum $G_0$,
which is here a definite indication of the Kondo effect.
Once $p\neq0$ the conductance becomes gradually suppressed at low frequencies,
and this process is more abrupt when the position of the OL
is tuned out of the symmetry point, as can be seen in Fig.~\ref{fig7}(b).
Moreover, this behavior is associated with the fact that the increasing of~$p$
essentially means that the spin-exchange processes responsible for the Kondo effect become less frequent.
In particular, such processes can be viewed as effectively reversing
the spin of an electron tunneling through the OL of a molecule.
Since a large value of the spin polarization~$p$ corresponds to the situation
of a large disproportion between numbers of majority and minority electrons,
one thus expects that the efficiency of the spin-exchange
processes dramatically drops down with increasing $p$.
Furthermore, it can be seen that the suppression of conductance
for finite $p$ is accompanied by the occurrence of a peak,
marked by a finely dashed lines in Figs.~\ref{fig7}(a)-(b),
which is nothing but the resonance A observed in previous figures.
Importantly, the position of this peak shifts now towards
higher frequencies for larger $p$.
As at present there is no intrinsic magnetic anisotropy ($D=0$),
it clearly proves that the underlying mechanism must involve the spintronic exchange fields,
which depend on the spin polarization of electrodes.

A completely different behavior of conductance as a function of $p$
can be observed for the AFM $J$-coupling, see Figs.~\ref{fig7}(c)-(d).
In such a case the low-frequency conductance is only weakly sensitive to a finite $p$,
while for $p=0$ conductance smoothly approaches zero at $\omega\rightarrow0$,
which signifies the two-stage Kondo effect, as discussed at the beginning of Sec.~\ref{sec:Large_spin}.
More precisely, for $p>0$, the presence of exchange fields results in suppression of the 
second stage of screening and the conductance reaches a small finite
non-universal value~\cite{Wojcik_PhysRevB.91.134422/2014}.
This low-frequency value
is larger in the case of $\varepsilon=-U/3$
compared to the particle-hole symmetry point since 
the dipolar exchange field is larger than the quadrupolar field.
For larger $\omega$ a kink similar to that marked as C
in Figs.~\ref{fig4}-\ref{fig6} develops,
and the dependence of its position on $p$
can be understood analogously as discussed above. 
Interestingly, both for the case of $\varepsilon=-U/2$
[Fig.~\ref{fig7}(c)] and $\varepsilon=-U/2$ [Fig.~\ref{fig7}(d)]
the main change in conductance when varying~$p$
occurs in the transient frequency range of $\TKqd\lesssim\omega\lesssim U$.
Then, in the case of $p=0$, the conductance shows an enhancement
for $\omega \approx \TKqd$ due to the first-stage Kondo effect,
which, however, quickly drops down for $\omega < \TKqd$
due to the second stage of screening.
When the spin polarization increases,
the local maximum in $G_{\rm P}(\omega)$
becomes suppressed due to the spin-splitting 
of the OL by the exchange field.

On the other hand, for large frequencies
the behavior of conductance hardly depends on $p$,
since $\omega$ is then much larger
than induced exchange fields.
Moreover, comparing with the FM $J$-coupling,
see Figs.~\ref{fig7}(a)-(b),
the conductance dependence is then
qualitatively the same in both 
the FM and AFM $J$-coupling cases.

\subsection{Signatures of dynamic spin accumulation}

\begin{figure}[t]
	\includegraphics[width=0.95\columnwidth]{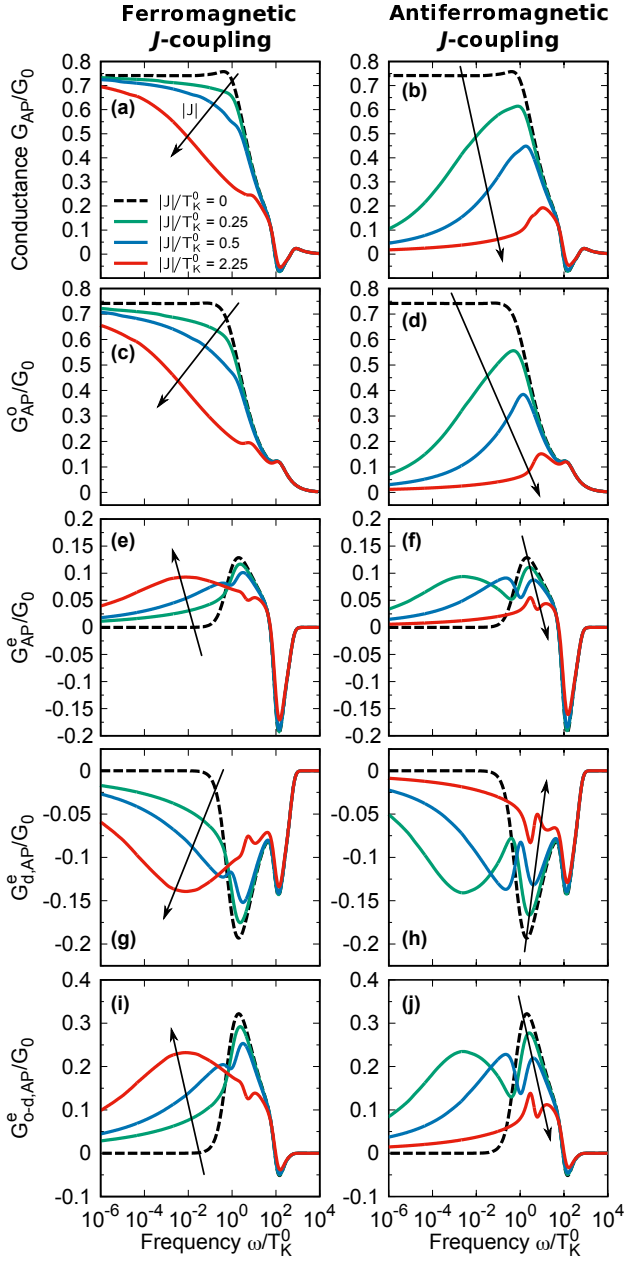}
	\caption{
	(color online)
	Decomposition of dynamical conductance (a)-(b) in the antiparallel magnetic configuration into dimensionless contributions from the odd (c)-(d) and even (e)-(j) channels shown for selected values of the exchange coupling~$J$ for $\varepsilon=-U/2$, $D=0$ and $p=0.5$.
	\emph{Left} (\emph{right}) \emph{column} corresponds to the FM (AFM) $J$-coupling case.
	Dashed lines represent the case of a QD ($J=0$), while black arrows indicate the direction of increasing $|J|$.
	Other parameters are the same as in Fig.~\ref{fig3}.
	\label{fig8}
	}
\end{figure}
\begin{figure}[t]
	\includegraphics[width=0.95\columnwidth]{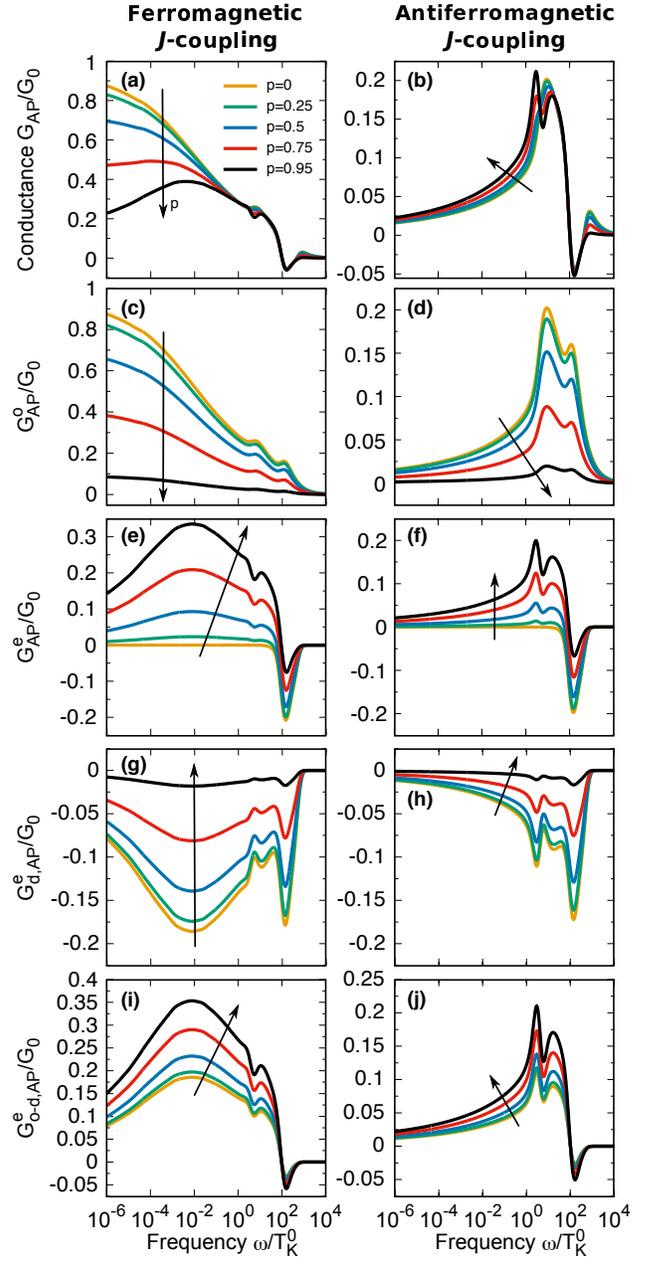}
	\caption{
	(color online)
	Analogous to Fig.~\ref{fig8} except that now conductance is decomposed for chosen values of the spin polarization parameter $p$ and $|J|/\TKqd=2.25$. Black arrows indicate here the direction of increasing $p$.
	\label{fig9}
	}
\end{figure}

In the last part of our work we address the effect of dynamical spin accumulation in the case of large-spin systems.
Such accumulation can be described by off-diagonal contribution of the 
frequency-dependent conductance, and corresponds to the situation when, e.g., one injects 
electrons of given spin orientation and detects the current of opposite spin direction.
This effect has recently attracted some attention in the case of quantum dots
\cite{Moca_Phys.Rev.B81/2010,Moca_Phys.Rev.B84/2011,Weymann_J.Appl.Phys.109/2011}.
It was predicted that the up-down component of the frequency-dependent
conductance is exclusively related to the even conduction channel,
it exhibits a maximum for $\omega \approx \TKqd$ and decays
as $\propto \omega^2$ with $\omega \to 0$.

To analyze and understand the effect of dynamical spin accumulation in the case of large-spin
molecules, let us focus on the situation where no effective exchange fields are present (antiparallel configuration),
and the intrinsic magnetic anisotropy is negligibly small ($D=0$).
In Fig.~\ref{fig8} we present the $\omega$-dependence of the total conductance,
$\GAP(\omega)$ (a,b), decomposed into contributions coming from the odd, $\GAP^o(\omega)$ (c,d) [see Eq.~(\ref{eq:Gc_odd})],
and even, $\GAP^e(\omega)$ (e,f) [see Eq.~(\ref{eq:Gc_even})], transport channels.
Additionally, the even conductance $\GAP^e(\omega)$ is further split into the diagonal, 
\begin{equation}
	G_\text{d,AP}^e(\omega)
	=
	\frac{1}{4}
	\sum_{\sigma}
	\Lambda_\sigma^\text{AP}
	\Upsilon_{\sigma\sigma}^\text{AP}
	\big[g_{\sigma\sigma}^e(\omega)\big]^{\!\text{AP}}		
	,
\end{equation} 
and off-diagonal,
\begin{equation}
	G_\text{o-d,AP}^e(\omega)
	=
	\frac{1}{4}
	\sum_{\sigma}
	\Lambda_\sigma^\text{AP}
	\Upsilon_{\sigma\overline{\sigma}}^\text{AP}
	\big[g_{\sigma\overline{\sigma}}^e(\omega)\big]^{\!\text{AP}}		
	,
\end{equation} 
parts, where the notation $\overline{\sigma}$ should be understood as $\overline{\uparrow}=\downarrow$ and $\overline{\downarrow}=\uparrow$. In fact, this is the off-diagonal part of the
even conductance that can be associated with dynamical spin accumulation in the system.
Importantly, the even contribution arises at finite frequencies
and vanishes when $\omega \to 0$.

An intuitive picture about how the effect of dynamical spin accumulation emerges in
a large-spin molecule can be gained by studying the evolution of the system from the case of a QD.
This can be achieved by a gradual increasing of the parameter $J$
from zero to a value used in previous sections.
This is illustrated in Fig.~\ref{fig8} where the dashed line represents the case of a QD~($J=0$).
One can immediately see that the conductance $\GAP^e(\omega)$ for a~QD 
in the even channel is enhanced at $\omega\approx\TKqd$,
which can be fully attributed to a sudden build-up of off-diagonal contribution,
see Fig.~\ref{fig8}(e).
Note that $\GAP^e(\omega)$, as well as its diagonal and off-diagonal components,
are very much suppressed until $\omega \gtrsim\TKqd$. 

This is, however, not the case for $J\neq0$,
where the even contribution affects the conductance at frequencies much smaller than $\TKqd$.
Moreover, the evolution of the even conductance, and thus the dynamical spin accumulation,
with increasing $|J|$ depends greatly on the type of exchange interaction $J$,
cf. the left and right column of Fig.~\ref{fig8}.
For the ferromagnetic exchange interaction,
with increasing $|J|$, a well-pronounced broad maximum in 
$G_\text{o-d,AP}^e(\omega)$ develops at frequencies
corresponding approximately to the Kondo temperature.
Note that~$\TK$ becomes suppressed with increasing $J$,
so that for large~$J$ the effect of dynamical spin accumulation
should be present at relatively low frequencies, see Fig.~\ref{fig8}(i).
On the other hand, a completely opposite behavior
can be observed in the AFM $J$-coupling case.
Now, the maximum spin accumulation arises at low frequencies for small~$|J|$,
and it shifts towards larger $\omega$ with increasing~$|J|$,
see Fig.~\ref{fig8}(j).
The frequency, at which this maximum occurs,
can be related to the energy scale ($\TK^*$) of the second stage
of Kondo screening, $\omega\approx \TK^*$.  Because this energy scale
becomes enhanced with increasing $|J|$, the position
of maximum in the even conductance and its off-diagonal contribution
moves toward larger $\omega$. Interestingly, 
$G_\text{o-d,AP}^e(\omega)$ exhibits then two peaks,
one occurring around $\omega \approx \TKqd$
and the other one for $\omega \approx \TK^*$, see Fig.~\ref{fig8}(j).
For very large exchange coupling,
$|J| \gtrsim \TKqd$, both the first and second stage of Kondo effect
become suppressed and so is the dynamical spin accumulation.
Summing up, while for the FM $J$-coupling the effect of the dynamical spin accumulation
becomes more and more important for large $|J|$ at small frequencies ($\omega\ll\TKqd$),
in the same limit it is gradually reduced for the AFM $J$-coupling,
so that for $|J|\gtrsim\TKqd$ this effect becomes significant only for $\omega\gtrsim\TKqd$.

Let us now investigate how the frequency-dependent conductance and
the effect of dynamical spin accumulation
are affected by the spin polarization $p$ of electrodes.
Figure~\ref{fig9} presents different contributions to the conductance
for a fixed value of $|J|$ and several values of $p$, as indicated in Fig.~\ref{fig9}(a). 
Because in the case of antiparallel configuration,
the effective molecule-electrode coupling is spin-independent,
the qualitative behavior of separate components of
odd and even conductance is the same as in the nonmagnetic systems.
Although the spin-dependence enters only through the coefficients
of the conductance, it can interestingly give rise to a highly nontrivial behavior.
This is, first of all, reflected in different dependence on $p$
of the contributions related to the spin diagonal and off-diagonal components.
As can be seen in Fig.~\ref{fig9}, increasing the spin polarization
results in a general suppression of the diagonal conductance (both odd and even),
while the off-diagonal component becomes enhanced regardless of the type of the $J$-coupling.
In fact, in the case of half-metallic leads, $p\to 1$,
the total conductance would be just given by 
$G_\text{o-d,AP}^e(\omega)$, i.e., the conductance
would be exclusively due to the effect of dynamical spin accumulation.
Thus, for large $p$, the frequency dependence of the total conductance
becomes mainly determined by the off-diagonal even channel
and characteristic features in $G_{\rm AP}(\omega)$ arise:
a broad maximum for $\omega\ll\TKqd$ in the case of the FM $J$-coupling,
see Fig.~\ref{fig9}(a),
and a sharp double-peak structure around $\omega\approx\TKqd$ for the AFM $J$-coupling,
see Fig.~\ref{fig9}(b).

Finally, we note that the high-frequency behavior of 
the conductance and its contributions
is essentially the same as in the case of quantum dots
---a large negative input due to the even channel,
which can be associated with dynamical charge accumulation,
is clearly visible, see Fig.~\ref{fig8}.
Moreover, we also notice that finite magnetic
anisotropy or the presence of exchange fields
(in parallel magnetic configuration)
generally results in the suppression of the dynamical spin accumulation.
Contrary to the case of quantum dots with ferromagnetic leads~\cite{Weymann_J.Appl.Phys.109/2011}, 
for large-spin molecules the spin accumulation can be also suppressed at the particle-hole symmetry point
due to the presence of the quadrupolar field.
The dynamical spin accumulation occurs then only for frequencies larger than
the relevant energy scales, corresponding to either magnetic anisotropy or exchange field,
whatever is larger.

\section{\label{sec:Conclusions}Conclusions}

In this paper we studied the frequency-dependent transport
through a large-spin magnetic molecule
coupled to ferromagnetic leads in the Kondo regime.
The molecule was modeled by an orbital level tunnel-coupled to the leads
and exchange-coupled to a core spin of the molecule.
By using the Kubo formula, we related the dynamical conductance
to correlation functions of the molecule and showed
that the conductance is generally given by two contributions:
the first one due to the odd linear combination of electron fields
in the left and right leads, and the second one, associated with the even combination.
While the odd conductance is responsible for low-frequency
conductance enhancement due to the Kondo effect,
the even conductance has a purely dynamical origin
and vanishes in the limit of $\omega \to 0$.
Using the numerical renormalization group method
to determine the relevant correlation functions, we studied
the dynamical properties of the system in the full parameter space,
in the case of both parallel and antiparallel magnetic configuration of the device.

We started with the case of a quantum dot as a reference
and performed a detailed analysis of frequency-dependent conductance of the considered magnetic molecule.
We showed that the behavior of the dynamical conductance depends
greatly on the type of molecule's internal exchange coupling and the magnetic configuration of the device.
In the case of antiparallel configuration, for antiferromagnetic exchange coupling,
the system exhibits the two-stage Kondo effect and $G_{\rm AP}(\omega)$
reveals a nonmonotonic dependence on $\omega$,
with a maximum around $\omega  \approx \TK$.
On the other hand, in the case of FM-$J$ coupling,
the system exhibits the underscreened Kondo effect
and the conductance increases at low frequencies due the Kondo effect.
We showed that the presence of finite magnetic anisotropy of the molecule,
generally suppresses the above-described behavior.

Interestingly, completely different features were observed in the case of parallel magnetic configuration.
First of all, in this case the spin-resolved couplings give rise to
dipolar and quadrupolar exchange fields, which 
have a strong influence on the behavior of dynamical conductance.
The dipolar field acts as an additional magnetic field,
while quadrupolar field induces
an additional effective anisotropy in the system.
The presence of those fields leads generally to the suppression of the Kondo effect
and, consequently, the frequency-dependent conductance.

Finally, we also analyzed the behavior of the off-diagonal
(in spin) component of the conductance,
which is responsible for dynamical spin accumulation in the molecule.
This contribution arises solely from
the even conduction channel and vanishes in the limit of $\omega\to 0$.
We found that the effect of dynamical spin accumulation
strongly depends on magnetic configuration of the system
and molecule's magnetic anisotropy. In particular, this effect
becomes especially pronounced for spin isotropic
molecules and in the case of antiparallel configuration.
Then, depending on the type of exchange interaction,
the off-diagonal conductance becomes enhanced
(suppressed) when the FM (AFM) $J$-coupling increases.
Moreover, we showed that for systems with large spin polarization of the leads,
the behavior of the total conductance is mainly determined
by the off-diagonal conductance and reveals 
a highly nontrivial behavior, with new local maxima
occurring at characteristic energy scales.
Then, the measurement of the total conductance
would allow for direct probing of dynamical spin accumulation in the system.

\acknowledgments

Work supported by the Polish Ministry of Science and Education
as Iuventus Plus project (IP2014 030973) in years 2015-2017 (A.P., M.M.)
and the Polish National Science Centre from funds awarded
through the decision No. DEC-2013/10/E/ST3/00213 (I.W.).
M.M. also acknowledges financial support from the Polish Ministry of Science
and Education through a young scientist fellowship (0066/E- 336/9/2014) and from the Knut and Alice Wallenberg Foundation.

\appendix
\section{\label{app:aux_deriv} Derivation of transport characteristics}

Below we provide the outline of derivation for some key formulae used in Sec.~\ref{sec:Dyn_cond}.
In particular, first, we obtain the general form for the frequency-dependent conductance $\acG_{\sigma\sigma^\prime}^{qq^\prime}(\omega)$, Eq.~(\ref{eq:G_omega_def}), and next we show how to calculate the Green's functions $\GF{\opI_\sigma^{q,e}}{\opI_{\sigma^\prime}^{q^\prime\!,e}}_\omega^\textrm{r}$, Eq.~(\ref{eq:GF_even}), and $\GF{\opI_\sigma^{q,o}}{\opI_{\sigma^\prime}^{q^\prime\!,o}}_\omega^\textrm{r}$, Eq.~(\ref{eq:GF_odd}), in terms of which $\acG_{\sigma\sigma^\prime}^{qq^\prime}(\omega)$ can be expressed, as shown in Eq.~(\ref{eq:GF_odd_even}).

\subsection*{Frequency-dependent conductance $\acG_{\sigma\sigma^\prime}^{qq^\prime}(\omega)$}

As explained in Sec.~\ref{sec:Dyn_cond}, within the linear-response transport regime the conductance $\acG_{\sigma\sigma^\prime}^{qq^\prime}(t-t^\prime)$ is essentially the retarded Green's function of the general form, see Eq.~(\ref{eq:G_vs_GF}),
\begin{equation}\label{eq:G_vs_GF_app}
	\acG_{\sigma\sigma^\prime}^{qq^\prime}(t-t^\prime)
	=
	-\frac{i}{\hbar}\theta(t-t^\prime)
	\big\langle[
	\opI_\sigma^q(t),
	\opQ_{\sigma^\prime}^{q^\prime}(t^\prime)
	]\big\rangle
	.
\end{equation}
In general, the frequency-dependent conductance $\acG_{\sigma\sigma^\prime}^{qq^\prime}(\omega)$ can be  derived from the expression above by considering the following equation of motion
\begin{equation}\label{eq:EOM}
	-
	\frac{\partial}{\partial \tau}
	\acG_{\sigma\sigma^\prime}^{qq^\prime}(\tau)
    =
    \GF{\opI_\sigma^q}{\opI_{\sigma^\prime}^{q^\prime}}_{\tau}^\text{r}
    +
    \frac{i}{\hbar} \delta (\tau) \big\langle[ \opI_\sigma^q(\tau),\opQ_{\sigma^\prime}^{q^\prime}(0)]\big\rangle
    ,
\end{equation}
which has been obtained from Eq.~(\ref{eq:G_vs_GF_app}) by differentiation with respect to $t^\prime$, and also the cyclic property of trace, $\text{Tr}\{\oper{\varrho}\ldots\}\equiv\langle\ldots\rangle$, has been used together with $\tau=t-t^\prime$. Additionally, in Eq.~(\ref{eq:EOM}) we have introduced the retarded Green's function for a current operator $\GF{\opI_\sigma^q}{\opI_{\sigma^\prime}^{q^\prime}}_{\tau}^\text{r}$ defined as
\begin{equation}
	\GF{\opI_\sigma^q}{\opI_{\sigma^\prime}^{q^\prime}}_{\tau}^\text{r}
	=
	-
	\frac{i}{\hbar}
	\theta(\tau)
	\big\langle \big[
	\opI_\sigma^q(\tau),
	\opI_{\sigma^\prime}^{q^\prime}(0)
	\big] \big\rangle
	.
\end{equation}
Next, we perform the Fourier transformation of the equation of motion~(\ref{eq:EOM})
\begin{align}
\hspace*{-1pt}
	-
	\int\!\!\intd \tau
	\Big[ \frac{\partial}{\partial\tau} \acG_{\sigma\sigma^\prime}^{qq^\prime}(\tau)\Big] \text{e}^{-i\omega\tau}
	=
	\int\!\!\intd \tau\,
	\GF{\opI_\sigma^q}{\opI_{\sigma^\prime}^{q^\prime}}_{\tau}^\text{r}
	\text{e}^{-i\omega\tau}
	&
	\nonumber\\
	+\
	\frac{i}{\hbar}
    \int\!\!\intd \tau\,
    \delta (\tau)
    \big\langle[ \opI_\sigma^q(\tau), \opQ_{\sigma^\prime}^{q^\prime}(0)]\big\rangle \text{e}^{-i\omega\tau}
    &
	,
\end{align}
which after some algebra leads to
\begin{equation}
	\acG_{\sigma\sigma^\prime}^{qq^\prime}(\omega)
	=
	\frac{i}{\omega}
	\Big[
	\GF{\opI_\sigma^q}{\opI_{\sigma^\prime}^{q^\prime}}_{\omega}^{r}
	+
	\frac{i}{\hbar}
	\big\langle[ \opI_\sigma^q(0) , \opQ_{\sigma^\prime}^{q^\prime}(0)]\big\rangle
    \Big]
    .
\end{equation}
Here, we note that $\big\langle[ \opI_\sigma^q(0) , \opQ_{\sigma^\prime}^{q^\prime}(0)]\big\rangle$ can be found from the equation above by considering the case of $\omega=0$, which yields
\begin{equation}
	\big\langle[ \opI_\sigma^q(0) , \opQ_{\sigma^\prime}^{q^\prime}(0)]\big\rangle
	=
    -\frac{i}{\hbar}
	\GF{\opI_\sigma^q}{\opI_{\sigma^\prime}^{q^\prime}}_{\omega=0}^{r}
	,
\end{equation}
so that the sought expression for the frequency-dependent (dynamical) conductance is finally found
\begin{equation}\label{eq:G_omega_def_app}
	\acG_{\sigma\sigma^\prime}^{qq^\prime}(\omega)
	=
	\frac{i}{\omega}
	\Big[\GF{\opI_\sigma^q}{\opI_{\sigma^\prime}^{q^\prime}}_\omega^\textrm{r}
	-
	\GF{\opI_\sigma^q}{\opI_{\sigma^\prime}^{q^\prime}}_{\omega=0}^\textrm{r}\Big].
\end{equation}
Since $\GF{\opI_\sigma^q}{\opI_{\sigma^\prime}^{q^\prime}}_{\omega=0}^\textrm{r}$ has to be real, one easily gets that
\begin{equation}
	\text{Re}\,\acG_{\sigma\sigma^\prime}^{qq^\prime}(\omega)
	=
	-\frac{1}{\omega}
	\text{Im}
	\GF{\opI_\sigma^q}{\opI_{\sigma^\prime}^{q^\prime}}_\omega^\textrm{r}
	,
\end{equation}
and
\begin{equation}
	\text{Im}\,\acG_{\sigma\sigma^\prime}^{qq^\prime}(\omega)
	=
	\frac{1}{\omega}
	\text{Re}
	\Big[\GF{\opI_\sigma^q}{\opI_{\sigma^\prime}^{q^\prime}}_\omega^\textrm{r}
	-
	\GF{\opI_\sigma^q}{\opI_{\sigma^\prime}^{q^\prime}}_{\omega=0}^\textrm{r}\Big]
	.
\end{equation}
In general, the frequency-dependent conductance $\acG_{\sigma\sigma^\prime}^{qq^\prime}(\omega)$, also referred to as admittance,  consists of real and imaginary parts. From the physics point of view,  the real part of admittance, commonly named \emph{conductance}, 
is a measure of the ease with which charge carriers can pass through the system.
The more easily charge carriers move, the higher is the conductance.
On the other hand, the imaginary part of admittance, so-called \emph{susceptance}, is the measure of how much a circuit opposes against conducting a time-dependent current.
Moreover, susceptance expresses the readiness
with which the system releases stored energy as the current and voltage fluctuate.

\subsection*{Retarded Green's functions for the even\\ and odd current operators}

As one can see from Eq.~(\ref{eq:G_omega_def_app}), calculation of the dynamical conductance  $\acG_{\sigma\sigma^\prime}^{qq^\prime}(\omega)$ basically means that one has to work out the Fourier transform of the retarded Green's function for the current operator~$\GF{\opI_\sigma^q}{\opI_{\sigma^\prime}^{q^\prime}}_\omega^\textrm{r}$. In Sec.~\ref{sec:Dyn_cond} we showed that this, in turn, can be further decomposed into two terms corresponding to currents in the `\emph{even}', $\GF{\opI_\sigma^{q,e}}{\opI_{\sigma^\prime}^{q^\prime\!,e}}_\omega^\textrm{r}$, and `\emph{odd}', $\GF{\opI_\sigma^{q,o}}{\opI_{\sigma^\prime}^{q^\prime\!,o}}_\omega^\textrm{r}$, channel, see Eq.~(\ref{eq:GF_odd_even}).
These two Green's functions can then be conveniently expressed in terms of the Green's functions for dimensionless current operators,
\begin{equation}\label{eq:GF_II_even_odd}
	\GF{\opII_\sigma^{e(o)}}{\opII_{\sigma^\prime}^{e(o)}}_\tau^\textrm{r}
	=
	-\frac{i}{\hbar}
	\theta(\tau)
	\big\langle[
	\opII_\sigma^{e(o)}(\tau),
	\opII_{\sigma^\prime}^{e(o)}(0)
	]\big\rangle
	,
\end{equation}
see Eq.~(\ref{eq:opIeo_def}) for definition of the operators $\opII_\sigma^{e(o)}$, as follows
\begin{equation}\label{eq:GF_even_t}
	\GF{\opI_\sigma^{q,e}}{\opI_{\sigma^\prime}^{q^\prime\!,e}}_\tau^\textrm{r}
	=
	-
	\frac{G_0}{\hbar\rho}
	\cdot
	\frac{\Gamma_\sigma^q\Gamma_{\sigma^\prime}^{q^\prime}}
	{\sqrt{\Gamma_\sigma^\text{eff}\Gamma_{\sigma^\prime}^\text{eff}}}
	\GF{\opII_\sigma^{e}}{\opII_{\sigma^\prime}^{e}}_\tau^\textrm{r}
	,
\end{equation}
with $G_0\equiv2e^2/h$ denoting the quantum of conductance, and
\begin{equation}\label{eq:GF_odd_t}
	\GF{\opI_\sigma^{q,o}}{\opI_{\sigma^\prime}^{q^\prime\!,o}}_t^\textrm{r}
	=
	-
	\eta_q\eta_{q^\prime}
	\frac{G_0}{\hbar\rho}
	\cdot
	\sqrt{
	\frac{\Gamma_\sigma^L\Gamma_\sigma^R
		\Gamma_{\sigma^\prime}^L
		\Gamma_{\sigma^\prime}^R}
	{\Gamma_\sigma^\text{eff}\Gamma_{\sigma^\prime}^\text{eff}}
	}
	\GF{\opII_\sigma^{o}}{\opII_{\sigma^\prime}^{o}}_\tau^\textrm{r}
	.
\end{equation}

It now goes without saying that $\GF{\opI_\sigma^{q,e}}{\opI_{\sigma^\prime}^{q^\prime\!,e}}_\omega^\textrm{r}$, Eq.~(\ref{eq:GF_even}), is straightforwardly obtained form Eq.~(\ref{eq:GF_even_t}) by means of the Fourier transform. On the other hand, an analogous procedure applied to Eq.~(\ref{eq:GF_odd_t}) is insufficient to validate Eq.~(\ref{eq:GF_odd}) for $\GF{\opI_\sigma^{q,o}}{\opI_{\sigma^\prime}^{q^\prime\!,o}}_\omega^\textrm{r}$. Importantly, one expects that the Fourier transform of the retarded Green's function for the current operator in the `\emph{odd}' channel $\GF{\opII_\sigma^{o}}{\opII_{\sigma^\prime}^{o}}_\omega^\textrm{r}$ can be further expressed in terms of the retarded  Green's functions for the OL $\GF{c_{\sigma}^{}}{c_{\sigma^\prime}^\dagger}_\omega^{\text{r}}$ and the equilibrium Fermi-Dirac distribution for electrodes $f(\omega)=\big\{1+\exp[\hbar\omega/(k_\text{B}T)]\big\}^{-1}$. To show this, let us insert the explicit form for the dimensionless current operator
$
	\opII_\sigma^o
	=
	\opc_\sigma^\dagger \hat{\Psi}_\sigma^o
	-
	\hat{\Psi}_\sigma^{o\dagger}\opc_\sigma^{}
$,
Eq.~(\ref{eq:opIIeo_def}), into Eq.~(\ref{eq:GF_II_even_odd}) and execute the commutator, which yields
\begin{align}\label{eq:GF_II_even_odd_2}
	\GF{\opII_\sigma^{q,o}}{\opII_{\sigma^\prime}^{q^\prime\!,o}}_\tau^\textrm{r}
	=
	i\hbar
	\Big[&
	\GF{\opc_{\sigma}^{}}{\opc_{\sigma^\prime}^{\dagger}}_\tau^\text{r}
	\GF{\hat{\Psi}_{\sigma^\prime}^o}{\hat{\Psi}_\sigma^{o\dagger}}_{-\tau}^<
	\nonumber\\[-2pt]
	+\ &
	\GF{\opc_{\sigma}^{}}{\opc_{\sigma^\prime}^{\dagger}}_\tau^<
	\GF{\hat{\Psi}_{\sigma^\prime}^o}{\hat{\Psi}_\sigma^{o\dagger}}_{-\tau}^\text{a}
	\nonumber\\
	+\ &
	\GF{\opc_{\sigma^\prime}^{}}{\opc_{\sigma}^{\dagger}}_{-\tau}^\text{a}
	\GF{\hat{\Psi}_{\sigma}^o}{\hat{\Psi}_{\sigma^\prime}^{o\dagger}}_\tau^<
	\nonumber\\
	+\ &
	\GF{\opc_{\sigma^\prime}^{}}{\opc_{\sigma}^{\dagger}}_{-\tau}^<
	\GF{\hat{\Psi}_{\sigma}^o}{\hat{\Psi}_{\sigma^\prime}^{o\dagger}}_\tau^\text{r}
	\Big]
	.
\end{align}
Here, we introduced a set of Green's functions for  fermionic operators $\opx_\sigma^{}=\opc_{\sigma}^{},\hat{\Psi}_{\sigma}^o$ defined as follows
\begin{equation}
	\left\{
	\begin{aligned}
	\GF{\opx_{\sigma}^{}}{\opx_{\sigma^\prime}^\dagger}_\tau^\text{r}
	=\ &
	-\frac{i}{\hbar}
	\theta(\tau)
	\big\langle
	\big\{\opx_\sigma^{}(\tau),\opx_{\sigma^\prime}^\dagger(0)\big\}
	\big\rangle
	,
	\\
	\GF{\opx_{\sigma}^{}}{\opx_{\sigma^\prime}^\dagger}_\tau^\text{a}
	=\ &
	\frac{i}{\hbar}
	\theta(-\tau)
	\big\langle
	\big\{\opx_\sigma^{}(\tau),\opx_{\sigma^\prime}^\dagger(0)\big\}
	\big\rangle
	,
	\\
	\GF{\opx_{\sigma}^{}}{\opx_{\sigma^\prime}^\dagger}_\tau^<
	=\ &
	\frac{i}{\hbar}
	\big\langle
	\opx_{\sigma^\prime}^\dagger(0)\opx_\sigma^{}(\tau)
	\big\rangle
	,
	\end{aligned}
	\right.
\end{equation}
and we also used the cyclic property of trace, that is, for instance $\big\langle\opx_\sigma^{}(0)\opx_{\sigma^\prime}^\dagger(\tau)\big\rangle=\big\langle\opx_\sigma^{}(-\tau)\opx_{\sigma^\prime}^\dagger(0)\big\rangle$.

In the next step, we perform the Fourier transformation of Eq.~(\ref{eq:GF_II_even_odd_2}), so that we derive
\begin{align}\label{eq:GF_II_even_odd_3}
	\GF{\opII_\sigma^{q,o}}{\opII_{\sigma^\prime}^{q^\prime\!,o}&}_\omega^\textrm{r}
	=
	\frac{i\hbar}{2\pi}
	\int\!\!\intd\tau
	\int\!\!\intd\omega^\prime
	\nonumber\\
	\Big[&
	\text{e}^{-i(\omega+\omega^\prime)\tau}
	\GF{\opc_{\sigma}^{}}{\opc_{\sigma^\prime}^{\dagger}}_\tau^\text{r}
	\GF{\hat{\Psi}_{\sigma^\prime}^o}{\hat{\Psi}_\sigma^{o\dagger}}_{\omega^\prime}^<
	\nonumber\\
	+\ &
	\text{e}^{-i(\omega+\omega^\prime)\tau}
	\GF{\opc_{\sigma}^{}}{\opc_{\sigma^\prime}^{\dagger}}_\tau^<
	\GF{\hat{\Psi}_{\sigma^\prime}^o}{\hat{\Psi}_\sigma^{o\dagger}}_{\omega^\prime}^\text{a}
	\nonumber\\
	+\ &
	\text{e}^{-i(\omega-\omega^\prime)\tau}
	\GF{\opc_{\sigma^\prime}^{}}{\opc_{\sigma}^{\dagger}}_{-\tau}^\text{a}
	\GF{\hat{\Psi}_{\sigma}^o}{\hat{\Psi}_{\sigma^\prime}^{o\dagger}}_{\omega^\prime}^<
	\nonumber\\
	+\ &
	\text{e}^{-i(\omega-\omega^\prime)\tau}
	\GF{\opc_{\sigma^\prime}^{}}{\opc_{\sigma}^{\dagger}}_{-\tau}^<
	\GF{\hat{\Psi}_{\sigma}^o}{\hat{\Psi}_{\sigma^\prime}^{o\dagger}}_{\omega^\prime}^\text{r}
	\Big]
	.
\end{align}
Now, recalling how the operator $\hat{\Psi}_\sigma^o=\sqrt{\rho}\int\intd\epsilon\,\opa_\sigma^o(\epsilon)$ is related to electrode operators $\opa_\sigma^L(\epsilon)$ and $\opa_\sigma^R(\epsilon)$, see Sec.~\ref{sec:Model}, and noting that electrodes are modeled as reservoirs of free electrons,  one finds~\cite{Bruus_book}:
$
	\GF{\hat{\Psi}_\sigma^o}{\hat{\Psi}_{\sigma^\prime}^{o\dagger}}_\omega^\text{r(a)}
	=
	\mp \delta_{\sigma\sigma^\prime}i\pi\rho
$
and
$
	\GF{\hat{\Psi}_\sigma^o}{\hat{\Psi}_{\sigma^\prime}^{o\dagger}}_\omega^<
	=
	\delta_{\sigma\sigma^\prime}
	i2\pi\rho f(\omega)
$.
Consequently, one can immediately see that after integrating Eq.~(\ref{eq:GF_II_even_odd_3}) with respect to $\tau$, the second and fourth term in the brackets cancel each other, so that the final result is reached
\begin{multline}
	\GF{\opII_\sigma^{q,o}}{\opII_{\sigma^\prime}^{q^\prime\!,o}}_\omega^\textrm{r}
	=
	-\delta_{\sigma\sigma^\prime}\hbar\rho
	\!
	\int\!\!\intd\omega^\prime
	\Big[
	f(\omega^\prime-\omega)
	\GF{\opc_{\sigma}^{}}{\opc_\sigma^{\dagger}}_{\omega^\prime}^\text{r}
	\\
	+
	f(\omega^\prime+\omega)
	\GF{\opc_\sigma^{}}{\opc_{\sigma}^{\dagger}}_{\omega^\prime}^\text{a}
	\Big]
	.
\end{multline}
To obtain Eq.~(\ref{eq:GF_odd}) from the expression above, one needs to insert it into the Fourier-transformed Eq.~(\ref{eq:GF_odd_t}) and use the property of the Green's functions, that is,
$
	\GF{\opc_\sigma^{}}{\opc_{\sigma}^{\dagger}}_{\omega^\prime}^\text{a}
	=
	\big[\GF{\opc_\sigma^{}}{\opc_{\sigma}^{\dagger}}_{\omega^\prime}^\text{r}\big]^\ast
$.
%


\begin{thebibliography}{68}%
\makeatletter
\providecommand \@ifxundefined [1]{%
 \@ifx{#1\undefined}
}%
\providecommand \@ifnum [1]{%
 \ifnum #1\expandafter \@firstoftwo
 \else \expandafter \@secondoftwo
 \fi
}%
\providecommand \@ifx [1]{%
 \ifx #1\expandafter \@firstoftwo
 \else \expandafter \@secondoftwo
 \fi
}%
\providecommand \natexlab [1]{#1}%
\providecommand \enquote  [1]{``#1''}%
\providecommand \bibnamefont  [1]{#1}%
\providecommand \bibfnamefont [1]{#1}%
\providecommand \citenamefont [1]{#1}%
\providecommand \href@noop [0]{\@secondoftwo}%
\providecommand \href [0]{\begingroup \@sanitize@url \@href}%
\providecommand \@href[1]{\@@startlink{#1}\@@href}%
\providecommand \@@href[1]{\endgroup#1\@@endlink}%
\providecommand \@sanitize@url [0]{\catcode `\\12\catcode `\$12\catcode
  `\&12\catcode `\#12\catcode `\^12\catcode `\_12\catcode `\%12\relax}%
\providecommand \@@startlink[1]{}%
\providecommand \@@endlink[0]{}%
\providecommand \url  [0]{\begingroup\@sanitize@url \@url }%
\providecommand \@url [1]{\endgroup\@href {#1}{\urlprefix }}%
\providecommand \urlprefix  [0]{URL }%
\providecommand \Eprint [0]{\href }%
\providecommand \doibase [0]{http://dx.doi.org/}%
\providecommand \selectlanguage [0]{\@gobble}%
\providecommand \bibinfo  [0]{\@secondoftwo}%
\providecommand \bibfield  [0]{\@secondoftwo}%
\providecommand \translation [1]{[#1]}%
\providecommand \BibitemOpen [0]{}%
\providecommand \bibitemStop [0]{}%
\providecommand \bibitemNoStop [0]{.\EOS\space}%
\providecommand \EOS [0]{\spacefactor3000\relax}%
\providecommand \BibitemShut  [1]{\csname bibitem#1\endcsname}%
\let\auto@bib@innerbib\@empty
\bibitem [{\citenamefont {Kahn}\ and\ \citenamefont
  {Martinez}(1998)}]{Kahn_Science279/1998}%
  \BibitemOpen
  \bibfield  {author} {\bibinfo {author} {\bibfnamefont {O.}~\bibnamefont
  {Kahn}}\ and\ \bibinfo {author} {\bibfnamefont {C.~J.}\ \bibnamefont
  {Martinez}},\ }\bibfield  {title} {\enquote {\bibinfo {title}
  {{Spin-transition polymers: from molecular materials toward memory
  devices}},}\ }\href@noop {} {\bibfield  {journal} {\bibinfo  {journal}
  {Science}\ }\textbf {\bibinfo {volume} {279}},\ \bibinfo {pages} {44}
  (\bibinfo {year} {1998})}\BibitemShut {NoStop}%
\bibitem [{\citenamefont {Leuenberger}\ and\ \citenamefont
  {Loss}(2001)}]{Leuenberger_Nature410/2001}%
  \BibitemOpen
  \bibfield  {author} {\bibinfo {author} {\bibfnamefont {M.~N.}\ \bibnamefont
  {Leuenberger}}\ and\ \bibinfo {author} {\bibfnamefont {D.}~\bibnamefont
  {Loss}},\ }\bibfield  {title} {\enquote {\bibinfo {title} {{Quantum computing
  in molecular magnets}},}\ }\href@noop {} {\bibfield  {journal} {\bibinfo
  {journal} {Nature}\ }\textbf {\bibinfo {volume} {410}},\ \bibinfo {pages}
  {789--793} (\bibinfo {year} {2001})}\BibitemShut {NoStop}%
\bibitem [{\citenamefont {Rocha}\ \emph {et~al.}(2005)\citenamefont {Rocha},
  \citenamefont {Garcia-Suarez}, \citenamefont {Bailey}, \citenamefont
  {Lambert}, \citenamefont {Ferrer},\ and\ \citenamefont
  {Sanvito}}]{Rocha_NatureMater.4/2005}%
  \BibitemOpen
  \bibfield  {author} {\bibinfo {author} {\bibfnamefont {A.~R.}\ \bibnamefont
  {Rocha}}, \bibinfo {author} {\bibfnamefont {V.~M.}\ \bibnamefont
  {Garcia-Suarez}}, \bibinfo {author} {\bibfnamefont {S.~W.}\ \bibnamefont
  {Bailey}}, \bibinfo {author} {\bibfnamefont {C.~J.}\ \bibnamefont {Lambert}},
  \bibinfo {author} {\bibfnamefont {J.}~\bibnamefont {Ferrer}}, \ and\ \bibinfo
  {author} {\bibfnamefont {S.}~\bibnamefont {Sanvito}},\ }\bibfield  {title}
  {\enquote {\bibinfo {title} {{Towards molecular spintronics}},}\ }\href@noop
  {} {\bibfield  {journal} {\bibinfo  {journal} {Nature Mater.}\ }\textbf
  {\bibinfo {volume} {4}},\ \bibinfo {pages} {335} (\bibinfo {year}
  {2005})}\BibitemShut {NoStop}%
\bibitem [{\citenamefont {Mannini}\ \emph {et~al.}(2009)\citenamefont
  {Mannini}, \citenamefont {Pineider}, \citenamefont {Sainctavit},
  \citenamefont {Danieli}, \citenamefont {Otero}, \citenamefont
  {Sciancalepore}, \citenamefont {Talarico}, \citenamefont {Arrio},
  \citenamefont {Cornia}, \citenamefont {Gatteschi},\ and\ \citenamefont
  {Sessoli}}]{Mannini_NatureMater.8/2009}%
  \BibitemOpen
  \bibfield  {author} {\bibinfo {author} {\bibfnamefont {M.}~\bibnamefont
  {Mannini}}, \bibinfo {author} {\bibfnamefont {F.}~\bibnamefont {Pineider}},
  \bibinfo {author} {\bibfnamefont {P.}~\bibnamefont {Sainctavit}}, \bibinfo
  {author} {\bibfnamefont {C.}~\bibnamefont {Danieli}}, \bibinfo {author}
  {\bibfnamefont {E.}~\bibnamefont {Otero}}, \bibinfo {author} {\bibfnamefont
  {C.}~\bibnamefont {Sciancalepore}}, \bibinfo {author} {\bibfnamefont {A.M.}\
  \bibnamefont {Talarico}}, \bibinfo {author} {\bibfnamefont {M.A.}\
  \bibnamefont {Arrio}}, \bibinfo {author} {\bibfnamefont {A.}~\bibnamefont
  {Cornia}}, \bibinfo {author} {\bibfnamefont {D.}~\bibnamefont {Gatteschi}}, \
  and\ \bibinfo {author} {\bibfnamefont {R.}~\bibnamefont {Sessoli}},\
  }\bibfield  {title} {\enquote {\bibinfo {title} {{Magnetic memory of a
  single-molecule quantum magnet wired to a gold surface}},}\ }\href@noop {}
  {\bibfield  {journal} {\bibinfo  {journal} {Nature Mater.}\ }\textbf
  {\bibinfo {volume} {8}},\ \bibinfo {pages} {194--197} (\bibinfo {year}
  {2009})}\BibitemShut {NoStop}%
\bibitem [{\citenamefont {Bartolom{\'e}}\ \emph {et~al.}(2014)\citenamefont
  {Bartolom{\'e}}, \citenamefont {Luis},\ and\ \citenamefont
  {Fern{\'a}ndez}}]{Bartolome_book}%
  \BibitemOpen
  \bibinfo {editor} {\bibfnamefont {J.}~\bibnamefont {Bartolom{\'e}}}, \bibinfo
  {editor} {\bibfnamefont {F.}~\bibnamefont {Luis}}, \ and\ \bibinfo {editor}
  {\bibfnamefont {J.~F.}\ \bibnamefont {Fern{\'a}ndez}},\ eds.,\ \href@noop {}
  {\emph {\bibinfo {title} {{Molecular magnets: Physics and AApplication}}}},\
  NanoScience and Technology\ (\bibinfo  {publisher} {Springer},\ \bibinfo
  {address} {Heidelberg},\ \bibinfo {year} {2014})\BibitemShut {NoStop}%
\bibitem [{\citenamefont {Sasaki}\ \emph {et~al.}(2000)\citenamefont {Sasaki},
  \citenamefont {De~Franceschi}, \citenamefont {Elzerman}, \citenamefont
  {Van~der Wiel}, \citenamefont {Eto}, \citenamefont {Tarucha},\ and\
  \citenamefont {Kouwenhoven}}]{Sasaki_Nature405/2000}%
  \BibitemOpen
  \bibfield  {author} {\bibinfo {author} {\bibfnamefont {S.}~\bibnamefont
  {Sasaki}}, \bibinfo {author} {\bibfnamefont {S.}~\bibnamefont
  {De~Franceschi}}, \bibinfo {author} {\bibfnamefont {J.M.}\ \bibnamefont
  {Elzerman}}, \bibinfo {author} {\bibfnamefont {W.G.}\ \bibnamefont {Van~der
  Wiel}}, \bibinfo {author} {\bibfnamefont {M.}~\bibnamefont {Eto}}, \bibinfo
  {author} {\bibfnamefont {S.}~\bibnamefont {Tarucha}}, \ and\ \bibinfo
  {author} {\bibfnamefont {L.P.}\ \bibnamefont {Kouwenhoven}},\ }\bibfield
  {title} {\enquote {\bibinfo {title} {Kondo effect in an integer-spin quantum
  dot},}\ }\href@noop {} {\bibfield  {journal} {\bibinfo  {journal} {Nature}\
  }\textbf {\bibinfo {volume} {405}},\ \bibinfo {pages} {764} (\bibinfo {year}
  {2000})}\BibitemShut {NoStop}%
\bibitem [{\citenamefont {Nyg{\aa}rd}\ \emph {et~al.}(2000)\citenamefont
  {Nyg{\aa}rd}, \citenamefont {Cobden},\ and\ \citenamefont
  {Lindelof}}]{Nygaard_Nature(London)408/2000}%
  \BibitemOpen
  \bibfield  {author} {\bibinfo {author} {\bibfnamefont {J.}~\bibnamefont
  {Nyg{\aa}rd}}, \bibinfo {author} {\bibfnamefont {D.H.}\ \bibnamefont
  {Cobden}}, \ and\ \bibinfo {author} {\bibfnamefont {P.E.}\ \bibnamefont
  {Lindelof}},\ }\bibfield  {title} {\enquote {\bibinfo {title} {{Kondo physics
  in carbon nanotubes}},}\ }\href@noop {} {\bibfield  {journal} {\bibinfo
  {journal} {Nature (London)}\ }\textbf {\bibinfo {volume} {408}},\ \bibinfo
  {pages} {342} (\bibinfo {year} {2000})}\BibitemShut {NoStop}%
\bibitem [{\citenamefont {Liang}\ \emph {et~al.}(2002)\citenamefont {Liang},
  \citenamefont {Shores}, \citenamefont {Bockrath}, \citenamefont {Long},\ and\
  \citenamefont {Park}}]{Liang_Nature417/2002}%
  \BibitemOpen
  \bibfield  {author} {\bibinfo {author} {\bibfnamefont {W.}~\bibnamefont
  {Liang}}, \bibinfo {author} {\bibfnamefont {M.P.}\ \bibnamefont {Shores}},
  \bibinfo {author} {\bibfnamefont {M.}~\bibnamefont {Bockrath}}, \bibinfo
  {author} {\bibfnamefont {J.R.}\ \bibnamefont {Long}}, \ and\ \bibinfo
  {author} {\bibfnamefont {H.}~\bibnamefont {Park}},\ }\bibfield  {title}
  {\enquote {\bibinfo {title} {{Kondo resonance in a single-molecule
  transistor}},}\ }\href@noop {} {\bibfield  {journal} {\bibinfo  {journal}
  {Nature}\ }\textbf {\bibinfo {volume} {417}},\ \bibinfo {pages} {725}
  (\bibinfo {year} {2002})}\BibitemShut {NoStop}%
\bibitem [{\citenamefont {Zhao}\ \emph {et~al.}(2005)\citenamefont {Zhao},
  \citenamefont {Li}, \citenamefont {Chen}, \citenamefont {Xiang},
  \citenamefont {Wang}, \citenamefont {Pan}, \citenamefont {Wang},
  \citenamefont {Xiao}, \citenamefont {Yang}, \citenamefont {Hou},\ and\
  \citenamefont {Zhu}}]{Zhao_Science309/2005}%
  \BibitemOpen
  \bibfield  {author} {\bibinfo {author} {\bibfnamefont {A.}~\bibnamefont
  {Zhao}}, \bibinfo {author} {\bibfnamefont {Q.}~\bibnamefont {Li}}, \bibinfo
  {author} {\bibfnamefont {L.}~\bibnamefont {Chen}}, \bibinfo {author}
  {\bibfnamefont {H.}~\bibnamefont {Xiang}}, \bibinfo {author} {\bibfnamefont
  {W.}~\bibnamefont {Wang}}, \bibinfo {author} {\bibfnamefont {S.}~\bibnamefont
  {Pan}}, \bibinfo {author} {\bibfnamefont {B.}~\bibnamefont {Wang}}, \bibinfo
  {author} {\bibfnamefont {X.}~\bibnamefont {Xiao}}, \bibinfo {author}
  {\bibfnamefont {J.}~\bibnamefont {Yang}}, \bibinfo {author} {\bibfnamefont
  {J.G.}\ \bibnamefont {Hou}}, \ and\ \bibinfo {author} {\bibfnamefont
  {Q.}~\bibnamefont {Zhu}},\ }\bibfield  {title} {\enquote {\bibinfo {title}
  {Controlling the kondo effect of an adsorbed magnetic ion through its
  chemical bonding},}\ }\href@noop {} {\bibfield  {journal} {\bibinfo
  {journal} {Science}\ }\textbf {\bibinfo {volume} {309}},\ \bibinfo {pages}
  {1542} (\bibinfo {year} {2005})}\BibitemShut {NoStop}%
\bibitem [{\citenamefont {Romeike}\ \emph
  {et~al.}(2006{\natexlab{a}})\citenamefont {Romeike}, \citenamefont
  {Wegewijs}, \citenamefont {Hofstetter},\ and\ \citenamefont
  {Schoeller}}]{Romeike_Phys.Rev.Lett.96/2006}%
  \BibitemOpen
  \bibfield  {author} {\bibinfo {author} {\bibfnamefont {C.}~\bibnamefont
  {Romeike}}, \bibinfo {author} {\bibfnamefont {M.~R.}\ \bibnamefont
  {Wegewijs}}, \bibinfo {author} {\bibfnamefont {W.}~\bibnamefont
  {Hofstetter}}, \ and\ \bibinfo {author} {\bibfnamefont {H.}~\bibnamefont
  {Schoeller}},\ }\bibfield  {title} {\enquote {\bibinfo {title}
  {{Quantum-tunneling-induced Kondo effect in single molecular magnets}},}\
  }\href@noop {} {\bibfield  {journal} {\bibinfo  {journal} {Phys. Rev. Lett.}\
  }\textbf {\bibinfo {volume} {96}},\ \bibinfo {pages} {196601} (\bibinfo
  {year} {2006}{\natexlab{a}})}\BibitemShut {NoStop}%
\bibitem [{\citenamefont {Romeike}\ \emph
  {et~al.}(2006{\natexlab{b}})\citenamefont {Romeike}, \citenamefont
  {Wegewijs}, \citenamefont {Hofstetter},\ and\ \citenamefont
  {Schoeller}}]{Romeike_Phys.Rev.Lett.97/2006}%
  \BibitemOpen
  \bibfield  {author} {\bibinfo {author} {\bibfnamefont {C.}~\bibnamefont
  {Romeike}}, \bibinfo {author} {\bibfnamefont {M.~R.}\ \bibnamefont
  {Wegewijs}}, \bibinfo {author} {\bibfnamefont {W.}~\bibnamefont
  {Hofstetter}}, \ and\ \bibinfo {author} {\bibfnamefont {H.}~\bibnamefont
  {Schoeller}},\ }\bibfield  {title} {\enquote {\bibinfo {title}
  {{Kondo-transport spectroscopy of single molecule magnets}},}\ }\href@noop {}
  {\bibfield  {journal} {\bibinfo  {journal} {Phys. Rev. Lett.}\ }\textbf
  {\bibinfo {volume} {97}},\ \bibinfo {pages} {206601} (\bibinfo {year}
  {2006}{\natexlab{b}})},\ \bibinfo {note} {see also: Phys. Rev. Lett
  \textbf{106}, 019902 (2011)}\BibitemShut {NoStop}%
\bibitem [{\citenamefont {Otte}\ \emph {et~al.}(2008)\citenamefont {Otte},
  \citenamefont {Ternes}, \citenamefont {von Bergmann}, \citenamefont {Loth},
  \citenamefont {Brune}, \citenamefont {Lutz}, \citenamefont {Hirjibehedin},\
  and\ \citenamefont {Heinrich}}]{Otte_NaturePhys.4/2008}%
  \BibitemOpen
  \bibfield  {author} {\bibinfo {author} {\bibfnamefont {A.F.}\ \bibnamefont
  {Otte}}, \bibinfo {author} {\bibfnamefont {M.}~\bibnamefont {Ternes}},
  \bibinfo {author} {\bibfnamefont {K.}~\bibnamefont {von Bergmann}}, \bibinfo
  {author} {\bibfnamefont {S.}~\bibnamefont {Loth}}, \bibinfo {author}
  {\bibfnamefont {H.}~\bibnamefont {Brune}}, \bibinfo {author} {\bibfnamefont
  {C.P.}\ \bibnamefont {Lutz}}, \bibinfo {author} {\bibfnamefont {C.F.}\
  \bibnamefont {Hirjibehedin}}, \ and\ \bibinfo {author} {\bibfnamefont {A.J.}\
  \bibnamefont {Heinrich}},\ }\bibfield  {title} {\enquote {\bibinfo {title}
  {{The role of magnetic anisotropy in the Kondo effect}},}\ }\href@noop {}
  {\bibfield  {journal} {\bibinfo  {journal} {Nature Phys.}\ }\textbf {\bibinfo
  {volume} {4}},\ \bibinfo {pages} {847--850} (\bibinfo {year}
  {2008})}\BibitemShut {NoStop}%
\bibitem [{\citenamefont {Roch}\ \emph {et~al.}(2008)\citenamefont {Roch},
  \citenamefont {Florens}, \citenamefont {Bouchiat}, \citenamefont
  {Wernsdorfer},\ and\ \citenamefont {Balestro}}]{Roch_Nature453/2008}%
  \BibitemOpen
  \bibfield  {author} {\bibinfo {author} {\bibfnamefont {N.}~\bibnamefont
  {Roch}}, \bibinfo {author} {\bibfnamefont {S.}~\bibnamefont {Florens}},
  \bibinfo {author} {\bibfnamefont {V.}~\bibnamefont {Bouchiat}}, \bibinfo
  {author} {\bibfnamefont {W.}~\bibnamefont {Wernsdorfer}}, \ and\ \bibinfo
  {author} {\bibfnamefont {F.}~\bibnamefont {Balestro}},\ }\bibfield  {title}
  {\enquote {\bibinfo {title} {{Quantum phase transition in a single-molecule
  quantum dot}},}\ }\href@noop {} {\bibfield  {journal} {\bibinfo  {journal}
  {Nature}\ }\textbf {\bibinfo {volume} {453}},\ \bibinfo {pages} {633}
  (\bibinfo {year} {2008})}\BibitemShut {NoStop}%
\bibitem [{\citenamefont {Scott}\ and\ \citenamefont
  {Natelson}(2010)}]{Scott_ACSNano4/2010}%
  \BibitemOpen
  \bibfield  {author} {\bibinfo {author} {\bibfnamefont {G.D.}\ \bibnamefont
  {Scott}}\ and\ \bibinfo {author} {\bibfnamefont {D.}~\bibnamefont
  {Natelson}},\ }\bibfield  {title} {\enquote {\bibinfo {title} {Kondo
  resonances in molecular devices},}\ }\href@noop {} {\bibfield  {journal}
  {\bibinfo  {journal} {ACS Nano}\ }\textbf {\bibinfo {volume} {4}},\ \bibinfo
  {pages} {3560} (\bibinfo {year} {2010})}\BibitemShut {NoStop}%
\bibitem [{\citenamefont {Parks}\ \emph {et~al.}(2010)\citenamefont {Parks},
  \citenamefont {Champagne}, \citenamefont {Costi}, \citenamefont {Shum},
  \citenamefont {Pasupathy}, \citenamefont {Neuscamman}, \citenamefont
  {Flores-Torres}, \citenamefont {Cornaglia}, \citenamefont {Aligia},
  \citenamefont {Balseiro}, \citenamefont {Chan}, \citenamefont {Abru\~{n}a},\
  and\ \citenamefont {Ralph}}]{Parks_Science328/2010}%
  \BibitemOpen
  \bibfield  {author} {\bibinfo {author} {\bibfnamefont {J.J.}\ \bibnamefont
  {Parks}}, \bibinfo {author} {\bibfnamefont {A.R.}\ \bibnamefont {Champagne}},
  \bibinfo {author} {\bibfnamefont {T.A.}\ \bibnamefont {Costi}}, \bibinfo
  {author} {\bibfnamefont {W.W.}\ \bibnamefont {Shum}}, \bibinfo {author}
  {\bibfnamefont {A.N.}\ \bibnamefont {Pasupathy}}, \bibinfo {author}
  {\bibfnamefont {E.}~\bibnamefont {Neuscamman}}, \bibinfo {author}
  {\bibfnamefont {S.}~\bibnamefont {Flores-Torres}}, \bibinfo {author}
  {\bibfnamefont {P.S.}\ \bibnamefont {Cornaglia}}, \bibinfo {author}
  {\bibfnamefont {A.A.}\ \bibnamefont {Aligia}}, \bibinfo {author}
  {\bibfnamefont {C.A.}\ \bibnamefont {Balseiro}}, \bibinfo {author}
  {\bibfnamefont {G.K.-L.}\ \bibnamefont {Chan}}, \bibinfo {author}
  {\bibfnamefont {H.D.}\ \bibnamefont {Abru\~{n}a}}, \ and\ \bibinfo {author}
  {\bibfnamefont {D.C.}\ \bibnamefont {Ralph}},\ }\bibfield  {title} {\enquote
  {\bibinfo {title} {{Mechanical control of spin states in spin-1 molecules and
  the underscreened Kondo effect}},}\ }\href@noop {} {\bibfield  {journal}
  {\bibinfo  {journal} {Science}\ }\textbf {\bibinfo {volume} {328}},\ \bibinfo
  {pages} {1370--1373} (\bibinfo {year} {2010})}\BibitemShut {NoStop}%
\bibitem [{\citenamefont {Grabert}\ and\ \citenamefont
  {Devoret}(1992)}]{Grabert_/1992}%
  \BibitemOpen
  \bibfield  {author} {\bibinfo {author} {\bibfnamefont {H.}~\bibnamefont
  {Grabert}}\ and\ \bibinfo {author} {\bibfnamefont {M.~H.}\ \bibnamefont
  {Devoret}},\ }\href@noop {} {\emph {\bibinfo {title} {{Single charge
  tunneling: Coulomb blockade phenomena in nanostructures}}}},\ NATO ASI Series
  B: Physics 294\ (\bibinfo  {publisher} {Plenum Press},\ \bibinfo {address}
  {New York},\ \bibinfo {year} {1992})\BibitemShut {NoStop}%
\bibitem [{\citenamefont {Kondo}(1964)}]{Kondo_Prog.Theor.Phys32/1964}%
  \BibitemOpen
  \bibfield  {author} {\bibinfo {author} {\bibfnamefont {J.}~\bibnamefont
  {Kondo}},\ }\bibfield  {title} {\enquote {\bibinfo {title} {{Resistance
  minimum in dilute magnetic alloys}},}\ }\href@noop {} {\bibfield  {journal}
  {\bibinfo  {journal} {Prog. Theor. Phys}\ }\textbf {\bibinfo {volume} {32}},\
  \bibinfo {pages} {37} (\bibinfo {year} {1964})}\BibitemShut {NoStop}%
\bibitem [{\citenamefont {Goldhaber-Gordon}\ \emph {et~al.}(1998)\citenamefont
  {Goldhaber-Gordon}, \citenamefont {Shtrikman}, \citenamefont {Mahalu},
  \citenamefont {Abusch-Magder}, \citenamefont {Meirav},\ and\ \citenamefont
  {Kastner}}]{Goldhaber_Nature391/98}%
  \BibitemOpen
  \bibfield  {author} {\bibinfo {author} {\bibfnamefont {D.}~\bibnamefont
  {Goldhaber-Gordon}}, \bibinfo {author} {\bibfnamefont {H.}~\bibnamefont
  {Shtrikman}}, \bibinfo {author} {\bibfnamefont {D.}~\bibnamefont {Mahalu}},
  \bibinfo {author} {\bibfnamefont {D.}~\bibnamefont {Abusch-Magder}}, \bibinfo
  {author} {\bibfnamefont {U.}~\bibnamefont {Meirav}}, \ and\ \bibinfo {author}
  {\bibfnamefont {M.A.}\ \bibnamefont {Kastner}},\ }\bibfield  {title}
  {\enquote {\bibinfo {title} {{The Kondo effect in a single-electron
  transistor}},}\ }\href@noop {} {\bibfield  {journal} {\bibinfo  {journal}
  {Nature (London)}\ }\textbf {\bibinfo {volume} {391}},\ \bibinfo {pages}
  {156--159} (\bibinfo {year} {1998})}\BibitemShut {NoStop}%
\bibitem [{\citenamefont {Cronenwett}\ \emph {et~al.}(1998)\citenamefont
  {Cronenwett}, \citenamefont {Oosterkamp},\ and\ \citenamefont
  {Kouwenhoven}}]{Cronenwett_Science281/1998}%
  \BibitemOpen
  \bibfield  {author} {\bibinfo {author} {\bibfnamefont {S.M.}\ \bibnamefont
  {Cronenwett}}, \bibinfo {author} {\bibfnamefont {T.H.}\ \bibnamefont
  {Oosterkamp}}, \ and\ \bibinfo {author} {\bibfnamefont {L.P.}\ \bibnamefont
  {Kouwenhoven}},\ }\bibfield  {title} {\enquote {\bibinfo {title} {{A tunable
  Kondo effect in quantum dots}},}\ }\href@noop {} {\bibfield  {journal}
  {\bibinfo  {journal} {Science}\ }\textbf {\bibinfo {volume} {281}},\ \bibinfo
  {pages} {540} (\bibinfo {year} {1998})}\BibitemShut {NoStop}%
\bibitem [{\citenamefont {Hewson}(1997)}]{Hewson_book}%
  \BibitemOpen
  \bibfield  {author} {\bibinfo {author} {\bibfnamefont {A.~C.}\ \bibnamefont
  {Hewson}},\ }\href@noop {} {\emph {\bibinfo {title} {{The Kondo problem to
  heavy fermions}}}}\ (\bibinfo  {publisher} {Cambridge University Press},\
  \bibinfo {address} {Cambridge},\ \bibinfo {year} {1997})\BibitemShut
  {NoStop}%
\bibitem [{\citenamefont {Koller}\ \emph {et~al.}(2005)\citenamefont {Koller},
  \citenamefont {Hewson},\ and\ \citenamefont
  {Meyer}}]{Koller_Phys.Rev.B72/2005}%
  \BibitemOpen
  \bibfield  {author} {\bibinfo {author} {\bibfnamefont {W.}~\bibnamefont
  {Koller}}, \bibinfo {author} {\bibfnamefont {A.~C.}\ \bibnamefont {Hewson}},
  \ and\ \bibinfo {author} {\bibfnamefont {D.}~\bibnamefont {Meyer}},\
  }\bibfield  {title} {\enquote {\bibinfo {title} {Singular dynamics of
  underscreened magnetic impurity models},}\ }\href@noop {} {\bibfield
  {journal} {\bibinfo  {journal} {Phys. Rev. B}\ }\textbf {\bibinfo {volume}
  {72}},\ \bibinfo {pages} {045117} (\bibinfo {year} {2005})}\BibitemShut
  {NoStop}%
\bibitem [{\citenamefont {Roch}\ \emph {et~al.}(2009)\citenamefont {Roch},
  \citenamefont {Florens}, \citenamefont {Costi}, \citenamefont {Wernsdorfer},\
  and\ \citenamefont {Balestro}}]{Roch_Phys.Rev.Lett.103/2009}%
  \BibitemOpen
  \bibfield  {author} {\bibinfo {author} {\bibfnamefont {N.}~\bibnamefont
  {Roch}}, \bibinfo {author} {\bibfnamefont {S.}~\bibnamefont {Florens}},
  \bibinfo {author} {\bibfnamefont {T.A.}\ \bibnamefont {Costi}}, \bibinfo
  {author} {\bibfnamefont {W.}~\bibnamefont {Wernsdorfer}}, \ and\ \bibinfo
  {author} {\bibfnamefont {F.}~\bibnamefont {Balestro}},\ }\bibfield  {title}
  {\enquote {\bibinfo {title} {{Observation of the underscreened Kondo effect
  in a molecular transistor}},}\ }\href@noop {} {\bibfield  {journal} {\bibinfo
   {journal} {Phys. Rev. Lett.}\ }\textbf {\bibinfo {volume} {103}},\ \bibinfo
  {pages} {197202} (\bibinfo {year} {2009})}\BibitemShut {NoStop}%
\bibitem [{\citenamefont {Weymann}\ and\ \citenamefont
  {Borda}(2010)}]{Weymann_Phys.Rev.B81/2010}%
  \BibitemOpen
  \bibfield  {author} {\bibinfo {author} {\bibfnamefont {I.}~\bibnamefont
  {Weymann}}\ and\ \bibinfo {author} {\bibfnamefont {L.}~\bibnamefont
  {Borda}},\ }\bibfield  {title} {\enquote {\bibinfo {title} {{Underscreened
  Kondo effect in quantum dots coupled to ferromagnetic leads}},}\ }\href@noop
  {} {\bibfield  {journal} {\bibinfo  {journal} {Phys. Rev. B}\ }\textbf
  {\bibinfo {volume} {81}},\ \bibinfo {pages} {115445} (\bibinfo {year}
  {2010})}\BibitemShut {NoStop}%
\bibitem [{\citenamefont {Cornaglia}\ \emph {et~al.}(2011)\citenamefont
  {Cornaglia}, \citenamefont {Roura~Bas}, \citenamefont {Aligia},\ and\
  \citenamefont {Balseiro}}]{Cornaglia_Europhys.Lett.93/2011}%
  \BibitemOpen
  \bibfield  {author} {\bibinfo {author} {\bibfnamefont {P.S.}\ \bibnamefont
  {Cornaglia}}, \bibinfo {author} {\bibfnamefont {P.}~\bibnamefont
  {Roura~Bas}}, \bibinfo {author} {\bibfnamefont {A.A.}\ \bibnamefont
  {Aligia}}, \ and\ \bibinfo {author} {\bibfnamefont {C.A.}\ \bibnamefont
  {Balseiro}},\ }\bibfield  {title} {\enquote {\bibinfo {title} {{Quantum
  transport through a stretched spin-1 molecule}},}\ }\href@noop {} {\bibfield
  {journal} {\bibinfo  {journal} {Europhys. Lett.}\ }\textbf {\bibinfo {volume}
  {93}},\ \bibinfo {pages} {47005} (\bibinfo {year} {2011})}\BibitemShut
  {NoStop}%
\bibitem [{\citenamefont {Misiorny}\ \emph
  {et~al.}(2012{\natexlab{a}})\citenamefont {Misiorny}, \citenamefont
  {Weymann},\ and\ \citenamefont {Barna{\'s}}}]{Misiorny_Phys.Rev.B86/2012_UK}%
  \BibitemOpen
  \bibfield  {author} {\bibinfo {author} {\bibfnamefont {M.}~\bibnamefont
  {Misiorny}}, \bibinfo {author} {\bibfnamefont {I.}~\bibnamefont {Weymann}}, \
  and\ \bibinfo {author} {\bibfnamefont {J.}~\bibnamefont {Barna{\'s}}},\
  }\bibfield  {title} {\enquote {\bibinfo {title} {{Underscreened Kondo effect
  in S= 1 magnetic quantum dots: Exchange, anisotropy, and temperature
  effects}},}\ }\href@noop {} {\bibfield  {journal} {\bibinfo  {journal} {Phys.
  Rev. B}\ }\textbf {\bibinfo {volume} {86}},\ \bibinfo {pages} {245415}
  (\bibinfo {year} {2012}{\natexlab{a}})}\BibitemShut {NoStop}%
\bibitem [{\citenamefont {van~der Wiel}\ \emph {et~al.}(2002)\citenamefont
  {van~der Wiel}, \citenamefont {De~Franceschi}, \citenamefont {Elzerman},
  \citenamefont {Tarucha}, \citenamefont {Kouwenhoven}, \citenamefont
  {Motohisa}, \citenamefont {Nakajima},\ and\ \citenamefont
  {Fukui}}]{Wiel_Phys.Rev.Lett.88/2002}%
  \BibitemOpen
  \bibfield  {author} {\bibinfo {author} {\bibfnamefont {W.G.}\ \bibnamefont
  {van~der Wiel}}, \bibinfo {author} {\bibfnamefont {S.}~\bibnamefont
  {De~Franceschi}}, \bibinfo {author} {\bibfnamefont {J.M.}\ \bibnamefont
  {Elzerman}}, \bibinfo {author} {\bibfnamefont {S.}~\bibnamefont {Tarucha}},
  \bibinfo {author} {\bibfnamefont {L.P.}\ \bibnamefont {Kouwenhoven}},
  \bibinfo {author} {\bibfnamefont {J.}~\bibnamefont {Motohisa}}, \bibinfo
  {author} {\bibfnamefont {F.}~\bibnamefont {Nakajima}}, \ and\ \bibinfo
  {author} {\bibfnamefont {T.}~\bibnamefont {Fukui}},\ }\bibfield  {title}
  {\enquote {\bibinfo {title} {Two-stage kondo effect in a quantum dot at a
  high magnetic field},}\ }\href@noop {} {\bibfield  {journal} {\bibinfo
  {journal} {Phys. Rev. Lett.}\ }\textbf {\bibinfo {volume} {88}},\ \bibinfo
  {pages} {126803} (\bibinfo {year} {2002})}\BibitemShut {NoStop}%
\bibitem [{\citenamefont {Posazhennikova}\ \emph {et~al.}(2007)\citenamefont
  {Posazhennikova}, \citenamefont {Bayani},\ and\ \citenamefont
  {Coleman}}]{Posazhennikova_Phys.Rev.B75/2007}%
  \BibitemOpen
  \bibfield  {author} {\bibinfo {author} {\bibfnamefont {A.}~\bibnamefont
  {Posazhennikova}}, \bibinfo {author} {\bibfnamefont {B.}~\bibnamefont
  {Bayani}}, \ and\ \bibinfo {author} {\bibfnamefont {P.}~\bibnamefont
  {Coleman}},\ }\bibfield  {title} {\enquote {\bibinfo {title} {{Conductance of
  a spin-1 quantum dot: The two-stage Kondo effect}},}\ }\href@noop {}
  {\bibfield  {journal} {\bibinfo  {journal} {Phys. Rev. B}\ }\textbf {\bibinfo
  {volume} {75}},\ \bibinfo {pages} {245329} (\bibinfo {year}
  {2007})}\BibitemShut {NoStop}%
\bibitem [{\citenamefont
  {{\v{Z}}itko}(2010)}]{Zitko_J.Phys.:Condens.Matter22/2010}%
  \BibitemOpen
  \bibfield  {author} {\bibinfo {author} {\bibfnamefont {R.}~\bibnamefont
  {{\v{Z}}itko}},\ }\bibfield  {title} {\enquote {\bibinfo {title} {Kondo
  screening in high-spin side-coupled two-impurity clusters},}\ }\href@noop {}
  {\bibfield  {journal} {\bibinfo  {journal} {J. Phys.: Condens. Matter}\
  }\textbf {\bibinfo {volume} {22}},\ \bibinfo {pages} {026002} (\bibinfo
  {year} {2010})}\BibitemShut {NoStop}%
\bibitem [{\citenamefont {Misiorny}\ \emph
  {et~al.}(2012{\natexlab{b}})\citenamefont {Misiorny}, \citenamefont
  {Weymann},\ and\ \citenamefont {Barna{\'s}}}]{Misiorny_Phys.Rev.B86/2012}%
  \BibitemOpen
  \bibfield  {author} {\bibinfo {author} {\bibfnamefont {M.}~\bibnamefont
  {Misiorny}}, \bibinfo {author} {\bibfnamefont {I.}~\bibnamefont {Weymann}}, \
  and\ \bibinfo {author} {\bibfnamefont {J.}~\bibnamefont {Barna{\'s}}},\
  }\bibfield  {title} {\enquote {\bibinfo {title} {Temperature dependence of
  electronic transport through molecular magnets in the kondo regime},}\
  }\href@noop {} {\bibfield  {journal} {\bibinfo  {journal} {Phys. Rev. B}\
  }\textbf {\bibinfo {volume} {86}},\ \bibinfo {pages} {035417} (\bibinfo
  {year} {2012}{\natexlab{b}})}\BibitemShut {NoStop}%
\bibitem [{\citenamefont {W\'ojcik}\ and\ \citenamefont
  {Weymann}(2015)}]{Wojcik_PhysRevB.91.134422/2014}%
  \BibitemOpen
  \bibfield  {author} {\bibinfo {author} {\bibfnamefont {K.~P.}\ \bibnamefont
  {W\'ojcik}}\ and\ \bibinfo {author} {\bibfnamefont {I.}~\bibnamefont
  {Weymann}},\ }\bibfield  {title} {\enquote {\bibinfo {title} {Two-stage kondo
  effect in t-shaped double quantum dots with ferromagnetic leads},}\
  }\href@noop {} {\bibfield  {journal} {\bibinfo  {journal} {Phys. Rev. B}\
  }\textbf {\bibinfo {volume} {91}},\ \bibinfo {pages} {134422} (\bibinfo
  {year} {2015})}\BibitemShut {NoStop}%
\bibitem [{\citenamefont {Misiorny}\ \emph
  {et~al.}(2011{\natexlab{a}})\citenamefont {Misiorny}, \citenamefont
  {Weymann},\ and\ \citenamefont {Barna\ifmmode~\acute{s}\else
  \'{s}\fi{}}}]{Misiorny_Phys.Rev.Lett.106/2011}%
  \BibitemOpen
  \bibfield  {author} {\bibinfo {author} {\bibfnamefont {M.}~\bibnamefont
  {Misiorny}}, \bibinfo {author} {\bibfnamefont {I.}~\bibnamefont {Weymann}}, \
  and\ \bibinfo {author} {\bibfnamefont {J.}~\bibnamefont
  {Barna\ifmmode~\acute{s}\else \'{s}\fi{}}},\ }\bibfield  {title} {\enquote
  {\bibinfo {title} {{Interplay of the Kondo effect and spin-polarized
  transport in magnetic molecules, adatoms, and quantum dots}},}\ }\href@noop
  {} {\bibfield  {journal} {\bibinfo  {journal} {Phys. Rev. Lett.}\ }\textbf
  {\bibinfo {volume} {106}},\ \bibinfo {pages} {126602} (\bibinfo {year}
  {2011}{\natexlab{a}})}\BibitemShut {NoStop}%
\bibitem [{\citenamefont {Misiorny}\ \emph
  {et~al.}(2011{\natexlab{b}})\citenamefont {Misiorny}, \citenamefont
  {Weymann},\ and\ \citenamefont {Barna\ifmmode~\acute{s}\else
  \'{s}\fi{}}}]{Misiorny_Phys.Rev.B84/2011}%
  \BibitemOpen
  \bibfield  {author} {\bibinfo {author} {\bibfnamefont {M.}~\bibnamefont
  {Misiorny}}, \bibinfo {author} {\bibfnamefont {I.}~\bibnamefont {Weymann}}, \
  and\ \bibinfo {author} {\bibfnamefont {J.}~\bibnamefont
  {Barna\ifmmode~\acute{s}\else \'{s}\fi{}}},\ }\bibfield  {title} {\enquote
  {\bibinfo {title} {{Influence of magnetic anisotropy on the Kondo effect and
  spin-polarized transport through magnetic molecules, adatoms, and quantum
  dots}},}\ }\href@noop {} {\bibfield  {journal} {\bibinfo  {journal} {Phys.
  Rev. B}\ }\textbf {\bibinfo {volume} {84}},\ \bibinfo {pages} {035445}
  (\bibinfo {year} {2011}{\natexlab{b}})}\BibitemShut {NoStop}%
\bibitem [{\citenamefont {Misiorny}\ and\ \citenamefont
  {Weymann}(2014)}]{Misiorny_Phys.Rev.B90/2014}%
  \BibitemOpen
  \bibfield  {author} {\bibinfo {author} {\bibfnamefont {M.}~\bibnamefont
  {Misiorny}}\ and\ \bibinfo {author} {\bibfnamefont {I.}~\bibnamefont
  {Weymann}},\ }\bibfield  {title} {\enquote {\bibinfo {title} {Transverse
  anisotropy effects on spin-resolved transport through large-spin
  molecules},}\ }\href@noop {} {\bibfield  {journal} {\bibinfo  {journal}
  {Phys. Rev. B}\ }\textbf {\bibinfo {volume} {90}},\ \bibinfo {pages} {235409}
  (\bibinfo {year} {2014})}\BibitemShut {NoStop}%
\bibitem [{\citenamefont {Misiorny}\ \emph {et~al.}(2013)\citenamefont
  {Misiorny}, \citenamefont {Hell},\ and\ \citenamefont
  {Wegewijs}}]{Misiorny_NaturePhys.9/2013}%
  \BibitemOpen
  \bibfield  {author} {\bibinfo {author} {\bibfnamefont {M.}~\bibnamefont
  {Misiorny}}, \bibinfo {author} {\bibfnamefont {M.}~\bibnamefont {Hell}}, \
  and\ \bibinfo {author} {\bibfnamefont {M.R.}\ \bibnamefont {Wegewijs}},\
  }\bibfield  {title} {\enquote {\bibinfo {title} {Spintronic magnetic
  anisotropy},}\ }\href@noop {} {\bibfield  {journal} {\bibinfo  {journal}
  {Nature Phys.}\ }\textbf {\bibinfo {volume} {9}},\ \bibinfo {pages}
  {801--805} (\bibinfo {year} {2013})}\BibitemShut {NoStop}%
\bibitem [{\citenamefont {Martinek}\ \emph
  {et~al.}(2003{\natexlab{a}})\citenamefont {Martinek}, \citenamefont {Utsumi},
  \citenamefont {Imamura}, \citenamefont {Barna\'{s}}, \citenamefont {Maekawa},
  \citenamefont {K\"onig},\ and\ \citenamefont
  {Sch\"on}}]{Martinek_Phys.Rev.Lett.91/2003_127203}%
  \BibitemOpen
  \bibfield  {author} {\bibinfo {author} {\bibfnamefont {J.}~\bibnamefont
  {Martinek}}, \bibinfo {author} {\bibfnamefont {Y.}~\bibnamefont {Utsumi}},
  \bibinfo {author} {\bibfnamefont {H.}~\bibnamefont {Imamura}}, \bibinfo
  {author} {\bibfnamefont {J.}~\bibnamefont {Barna\'{s}}}, \bibinfo {author}
  {\bibfnamefont {S.}~\bibnamefont {Maekawa}}, \bibinfo {author} {\bibfnamefont
  {J.}~\bibnamefont {K\"onig}}, \ and\ \bibinfo {author} {\bibfnamefont
  {G.}~\bibnamefont {Sch\"on}},\ }\bibfield  {title} {\enquote {\bibinfo
  {title} {{Kondo effect in quantum dots coupled to ferromagnetic leads}},}\
  }\href@noop {} {\bibfield  {journal} {\bibinfo  {journal} {Phys. Rev. Lett.}\
  }\textbf {\bibinfo {volume} {91}},\ \bibinfo {pages} {127203} (\bibinfo
  {year} {2003}{\natexlab{a}})}\BibitemShut {NoStop}%
\bibitem [{\citenamefont {Heikkil{\"a}}(2013)}]{Heikkilae_book}%
  \BibitemOpen
  \bibfield  {author} {\bibinfo {author} {\bibfnamefont {T.~T.}\ \bibnamefont
  {Heikkil{\"a}}},\ }\href@noop {} {\emph {\bibinfo {title} {{The physics of
  nanoelectronics: Transport and fluctuation phenomena at low
  temperatures}}}},\ \bibinfo {series} {Oxford Master Series in Physics},
  Vol.~\bibinfo {volume} {21}\ (\bibinfo  {publisher} {OUP},\ \bibinfo
  {address} {Oxford},\ \bibinfo {year} {2013})\BibitemShut {NoStop}%
\bibitem [{\citenamefont {Kubo}\ \emph {et~al.}(1986)\citenamefont {Kubo},
  \citenamefont {Toda},\ and\ \citenamefont {Hashitsume}}]{Kubo_book}%
  \BibitemOpen
  \bibfield  {author} {\bibinfo {author} {\bibfnamefont {R.}~\bibnamefont
  {Kubo}}, \bibinfo {author} {\bibfnamefont {M.}~\bibnamefont {Toda}}, \ and\
  \bibinfo {author} {\bibfnamefont {N.}~\bibnamefont {Hashitsume}},\
  }\href@noop {} {\emph {\bibinfo {title} {{Statistical physics II:
  Nonequilibrium statistical mechanics}}}},\ \bibinfo {edition} {1st}\ ed.,\
  \bibinfo {series} {{Springer Series in Solid-State Sciences}}, Vol.~\bibinfo
  {volume} {31}\ (\bibinfo  {publisher} {Springer},\ \bibinfo {address}
  {Heidelberg},\ \bibinfo {year} {1986})\BibitemShut {NoStop}%
\bibitem [{\citenamefont {Sindel}\ \emph {et~al.}(2005)\citenamefont {Sindel},
  \citenamefont {Hofstetter}, \citenamefont {Von~Delft},\ and\ \citenamefont
  {Kindermann}}]{Sindel_Phys.Rev.Lett.94/2005}%
  \BibitemOpen
  \bibfield  {author} {\bibinfo {author} {\bibfnamefont {M.}~\bibnamefont
  {Sindel}}, \bibinfo {author} {\bibfnamefont {W.}~\bibnamefont {Hofstetter}},
  \bibinfo {author} {\bibfnamefont {J.}~\bibnamefont {Von~Delft}}, \ and\
  \bibinfo {author} {\bibfnamefont {M.}~\bibnamefont {Kindermann}},\ }\bibfield
   {title} {\enquote {\bibinfo {title} {{Frequency-dependent transport through
  a quantum dot in the Kondo regime}},}\ }\href@noop {} {\bibfield  {journal}
  {\bibinfo  {journal} {Phys. Rev. Lett.}\ }\textbf {\bibinfo {volume} {94}},\
  \bibinfo {pages} {196602} (\bibinfo {year} {2005})}\BibitemShut {NoStop}%
\bibitem [{\citenamefont {T{\'o}th}\ \emph {et~al.}(2007)\citenamefont
  {T{\'o}th}, \citenamefont {Borda}, \citenamefont {von Delft},\ and\
  \citenamefont {Zar{\'a}nd}}]{Toth_Phys.Rev.B76/2007}%
  \BibitemOpen
  \bibfield  {author} {\bibinfo {author} {\bibfnamefont {A.I.}\ \bibnamefont
  {T{\'o}th}}, \bibinfo {author} {\bibfnamefont {L.}~\bibnamefont {Borda}},
  \bibinfo {author} {\bibfnamefont {J.}~\bibnamefont {von Delft}}, \ and\
  \bibinfo {author} {\bibfnamefont {G}~\bibnamefont {Zar{\'a}nd}},\ }\bibfield
  {title} {\enquote {\bibinfo {title} {{Dynamical conductance in the
  two-channel Kondo regime of a double dot system}},}\ }\href@noop {}
  {\bibfield  {journal} {\bibinfo  {journal} {Phys. Rev. B}\ }\textbf {\bibinfo
  {volume} {76}},\ \bibinfo {pages} {155318} (\bibinfo {year}
  {2007})}\BibitemShut {NoStop}%
\bibitem [{\citenamefont {Moca}\ \emph {et~al.}(2010)\citenamefont {Moca},
  \citenamefont {Weymann},\ and\ \citenamefont
  {Zar{\'a}nd}}]{Moca_Phys.Rev.B81/2010}%
  \BibitemOpen
  \bibfield  {author} {\bibinfo {author} {\bibfnamefont {C.P.}\ \bibnamefont
  {Moca}}, \bibinfo {author} {\bibfnamefont {I.}~\bibnamefont {Weymann}}, \
  and\ \bibinfo {author} {\bibfnamefont {G.}~\bibnamefont {Zar{\'a}nd}},\
  }\bibfield  {title} {\enquote {\bibinfo {title} {Theory of
  frequency-dependent spin current noise through correlated quantum dots},}\
  }\href@noop {} {\bibfield  {journal} {\bibinfo  {journal} {Phys. Rev. B}\
  }\textbf {\bibinfo {volume} {81}},\ \bibinfo {pages} {241305} (\bibinfo
  {year} {2010})}\BibitemShut {NoStop}%
\bibitem [{\citenamefont {Moca}\ \emph
  {et~al.}(2011{\natexlab{a}})\citenamefont {Moca}, \citenamefont {Weymann},\
  and\ \citenamefont {Zarand}}]{Moca_Phys.Rev.B84/2011}%
  \BibitemOpen
  \bibfield  {author} {\bibinfo {author} {\bibfnamefont {C.P.}\ \bibnamefont
  {Moca}}, \bibinfo {author} {\bibfnamefont {I.}~\bibnamefont {Weymann}}, \
  and\ \bibinfo {author} {\bibfnamefont {G.}~\bibnamefont {Zarand}},\
  }\bibfield  {title} {\enquote {\bibinfo {title} {{Theory of ac spin current
  noise and spin conductance through a quantum dot in the Kondo regime:
  Equilibrium case}},}\ }\href@noop {} {\bibfield  {journal} {\bibinfo
  {journal} {Phys. Rev. B}\ }\textbf {\bibinfo {volume} {84}},\ \bibinfo
  {pages} {235441} (\bibinfo {year} {2011}{\natexlab{a}})}\BibitemShut
  {NoStop}%
\bibitem [{\citenamefont {Moca}\ \emph
  {et~al.}(2011{\natexlab{b}})\citenamefont {Moca}, \citenamefont {Simon},
  \citenamefont {Chung},\ and\ \citenamefont
  {Zar{\'a}nd}}]{Moca_Phys.Rev.B83/2011}%
  \BibitemOpen
  \bibfield  {author} {\bibinfo {author} {\bibfnamefont {C.P.}\ \bibnamefont
  {Moca}}, \bibinfo {author} {\bibfnamefont {P.}~\bibnamefont {Simon}},
  \bibinfo {author} {\bibfnamefont {C.H.}\ \bibnamefont {Chung}}, \ and\
  \bibinfo {author} {\bibfnamefont {G.}~\bibnamefont {Zar{\'a}nd}},\ }\bibfield
   {title} {\enquote {\bibinfo {title} {{Nonequilibrium frequency-dependent
  noise through a quantum dot: A real-time functional renormalization group
  approach}},}\ }\href@noop {} {\bibfield  {journal} {\bibinfo  {journal}
  {Phys. Rev. B}\ }\textbf {\bibinfo {volume} {83}},\ \bibinfo {pages} {201303}
  (\bibinfo {year} {2011}{\natexlab{b}})}\BibitemShut {NoStop}%
\bibitem [{\citenamefont {Weymann}\ and\ \citenamefont
  {Moca}(2011)}]{Weymann_J.Appl.Phys.109/2011}%
  \BibitemOpen
  \bibfield  {author} {\bibinfo {author} {\bibfnamefont {I.}~\bibnamefont
  {Weymann}}\ and\ \bibinfo {author} {\bibfnamefont {C.P.}\ \bibnamefont
  {Moca}},\ }\bibfield  {title} {\enquote {\bibinfo {title}
  {{Frequency-dependent conductance of Kondo quantum dots coupled to
  ferromagnetic leads}},}\ }\href@noop {} {\bibfield  {journal} {\bibinfo
  {journal} {J. Appl. Phys.}\ }\textbf {\bibinfo {volume} {109}},\ \bibinfo
  {pages} {07C704} (\bibinfo {year} {2011})}\BibitemShut {NoStop}%
\bibitem [{\citenamefont {Moca}\ \emph {et~al.}(2014)\citenamefont {Moca},
  \citenamefont {Simon}, \citenamefont {Chung},\ and\ \citenamefont
  {Zarand}}]{Moca_Phys.Rev.B89/2014}%
  \BibitemOpen
  \bibfield  {author} {\bibinfo {author} {\bibfnamefont {C.P.}\ \bibnamefont
  {Moca}}, \bibinfo {author} {\bibfnamefont {P.}~\bibnamefont {Simon}},
  \bibinfo {author} {\bibfnamefont {C.-H.}\ \bibnamefont {Chung}}, \ and\
  \bibinfo {author} {\bibfnamefont {G.}~\bibnamefont {Zarand}},\ }\bibfield
  {title} {\enquote {\bibinfo {title} {Finite-frequency-dependent noise of a
  quantum dot in a magnetic field},}\ }\href@noop {} {\bibfield  {journal}
  {\bibinfo  {journal} {Phys. Rev. B}\ }\textbf {\bibinfo {volume} {89}},\
  \bibinfo {pages} {155138} (\bibinfo {year} {2014})}\BibitemShut {NoStop}%
\bibitem [{\citenamefont {E.~Zakka-Bajjani}\ and\ \citenamefont
  {Jin}(2007)}]{Zakka_PRL.99/2007}%
  \BibitemOpen
  \bibfield  {author} {\bibinfo {author} {\bibfnamefont {F.~Portier P. Roche D.
  C. Glattli A.~Cavanna}\ \bibnamefont {E.~Zakka-Bajjani}, \bibfnamefont
  {J.~Segala}}\ and\ \bibinfo {author} {\bibfnamefont {Y.}~\bibnamefont
  {Jin}},\ }\bibfield  {title} {\enquote {\bibinfo {title} {{Experimental test
  of the high-frequency quantum shot noise theory in a quantum point
  contact}},}\ }\href@noop {} {\bibfield  {journal} {\bibinfo  {journal} {Phys.
  Rev. Lett.}\ }\textbf {\bibinfo {volume} {99}},\ \bibinfo {pages} {236803}
  (\bibinfo {year} {2007})}\BibitemShut {NoStop}%
\bibitem [{\citenamefont {Gabelli}\ and\ \citenamefont
  {Reulet}(2008)}]{Gabelli_PRL.100/2008}%
  \BibitemOpen
  \bibfield  {author} {\bibinfo {author} {\bibfnamefont {J.}~\bibnamefont
  {Gabelli}}\ and\ \bibinfo {author} {\bibfnamefont {B.}~\bibnamefont
  {Reulet}},\ }\bibfield  {title} {\enquote {\bibinfo {title} {{ Dynamics of
  quantum noise in a tunnel junction under ac excitation }},}\ }\href@noop {}
  {\bibfield  {journal} {\bibinfo  {journal} {Phys. Rev. Lett.}\ }\textbf
  {\bibinfo {volume} {100}},\ \bibinfo {pages} {026601} (\bibinfo {year}
  {2008})}\BibitemShut {NoStop}%
\bibitem [{\citenamefont {E.~Zakka-Bajjani}\ and\ \citenamefont
  {Portier}(2010)}]{Zakka_PRL.104/206802}%
  \BibitemOpen
  \bibfield  {author} {\bibinfo {author} {\bibfnamefont {N.~Coulombel P. Roche
  C. D.~Glattli}\ \bibnamefont {E.~Zakka-Bajjani}, \bibfnamefont
  {J.~Dufouleur}}\ and\ \bibinfo {author} {\bibfnamefont {F.}~\bibnamefont
  {Portier}},\ }\bibfield  {title} {\enquote {\bibinfo {title} {{ Experimental
  determination of the statistics of photons emitted by a tunnel junction}},}\
  }\href@noop {} {\bibfield  {journal} {\bibinfo  {journal} {Phys. Rev. Lett.}\
  }\textbf {\bibinfo {volume} {104}},\ \bibinfo {pages} {206802} (\bibinfo
  {year} {2010})}\BibitemShut {NoStop}%
\bibitem [{\citenamefont {J.~Basset}\ and\ \citenamefont
  {Deblock}(2010)}]{Basset_PRL.105/166801}%
  \BibitemOpen
  \bibfield  {author} {\bibinfo {author} {\bibfnamefont {H.~Bouchiat}\
  \bibnamefont {J.~Basset}}\ and\ \bibinfo {author} {\bibfnamefont
  {R.}~\bibnamefont {Deblock}},\ }\bibfield  {title} {\enquote {\bibinfo
  {title} {{Emission and absorption quantum noise measurement with an on-chip
  resonant circuit}},}\ }\href@noop {} {\bibfield  {journal} {\bibinfo
  {journal} {Phys. Rev. Lett.}\ }\textbf {\bibinfo {volume} {105}},\ \bibinfo
  {pages} {166801} (\bibinfo {year} {2010})}\BibitemShut {NoStop}%
\bibitem [{\citenamefont {Basset}\ \emph {et~al.}(2012)\citenamefont {Basset},
  \citenamefont {Kasumov}, \citenamefont {Moca}, \citenamefont {Zar\'and},
  \citenamefont {Simon}, \citenamefont {Bouchiat},\ and\ \citenamefont
  {Deblock}}]{Basset_PRL.108/046802}%
  \BibitemOpen
  \bibfield  {author} {\bibinfo {author} {\bibfnamefont {J.}~\bibnamefont
  {Basset}}, \bibinfo {author} {\bibfnamefont {A.~Yu.}\ \bibnamefont
  {Kasumov}}, \bibinfo {author} {\bibfnamefont {C.~P.}\ \bibnamefont {Moca}},
  \bibinfo {author} {\bibfnamefont {G.}~\bibnamefont {Zar\'and}}, \bibinfo
  {author} {\bibfnamefont {P.}~\bibnamefont {Simon}}, \bibinfo {author}
  {\bibfnamefont {H.}~\bibnamefont {Bouchiat}}, \ and\ \bibinfo {author}
  {\bibfnamefont {R.}~\bibnamefont {Deblock}},\ }\bibfield  {title} {\enquote
  {\bibinfo {title} {Measurement of quantum noise in a carbon nanotube quantum
  dot in the kondo regime},}\ }\href@noop {} {\bibfield  {journal} {\bibinfo
  {journal} {Phys. Rev. Lett.}\ }\textbf {\bibinfo {volume} {108}},\ \bibinfo
  {pages} {046802} (\bibinfo {year} {2012})}\BibitemShut {NoStop}%
\bibitem [{\citenamefont {Wilson}(1975)}]{Wilson_RMP.47/773}%
  \BibitemOpen
  \bibfield  {author} {\bibinfo {author} {\bibfnamefont {K.~G.}\ \bibnamefont
  {Wilson}},\ }\bibfield  {title} {\enquote {\bibinfo {title} {{The
  renormalization group: critical phenomena and the Kondo problem.}}}\
  }\href@noop {} {\bibfield  {journal} {\bibinfo  {journal} {Rev. Mod. Phys.}\
  }\textbf {\bibinfo {volume} {47}},\ \bibinfo {pages} {773} (\bibinfo {year}
  {1975})}\BibitemShut {NoStop}%
\bibitem [{\citenamefont {Glazman}\ and\ \citenamefont
  {Raikh}(1988)}]{Glazman_JETP.Lett.47/1988}%
  \BibitemOpen
  \bibfield  {author} {\bibinfo {author} {\bibfnamefont {L.~I.}\ \bibnamefont
  {Glazman}}\ and\ \bibinfo {author} {\bibfnamefont {M.~E.}\ \bibnamefont
  {Raikh}},\ }\bibfield  {title} {\enquote {\bibinfo {title} {{Resonant Kondo
  transparency of a barrier with quasilocal impurity states}},}\ }\href@noop {}
  {\bibfield  {journal} {\bibinfo  {journal} {JETP. Lett.}\ }\textbf {\bibinfo
  {volume} {47}},\ \bibinfo {pages} {452--455} (\bibinfo {year}
  {1988})}\BibitemShut {NoStop}%
\bibitem [{\citenamefont {Bruus}\ and\ \citenamefont
  {Flensberg}(2004)}]{Bruus_book}%
  \BibitemOpen
  \bibfield  {author} {\bibinfo {author} {\bibfnamefont {H.}~\bibnamefont
  {Bruus}}\ and\ \bibinfo {author} {\bibfnamefont {K.}~\bibnamefont
  {Flensberg}},\ }\href@noop {} {\emph {\bibinfo {title} {{Many-body quantum
  theory in condesed matter physics: An introduction}}}},\ Oxford Graduate
  Texts\ (\bibinfo  {publisher} {OUP},\ \bibinfo {address} {Oxford},\ \bibinfo
  {year} {2004})\BibitemShut {NoStop}%
\bibitem [{\citenamefont {R.~Bulla}\ and\ \citenamefont
  {Pruschke}(2008)}]{Bulla_RMP.80/395}%
  \BibitemOpen
  \bibfield  {author} {\bibinfo {author} {\bibfnamefont {T.~A.~Costi}\
  \bibnamefont {R.~Bulla}}\ and\ \bibinfo {author} {\bibfnamefont
  {T.}~\bibnamefont {Pruschke}},\ }\bibfield  {title} {\enquote {\bibinfo
  {title} {{Numerical renormalization group method for quantum impurity
  systems.}}}\ }\href@noop {} {\bibfield  {journal} {\bibinfo  {journal} {Rev.
  Mod. Phys.}\ }\textbf {\bibinfo {volume} {80}},\ \bibinfo {pages} {395}
  (\bibinfo {year} {2008})}\BibitemShut {NoStop}%
\bibitem [{\citenamefont {Legeza}\ \emph {et~al.}(2008)\citenamefont {Legeza},
  \citenamefont {Moca}, \citenamefont {T\'{o}th}, \citenamefont {Weymann},\
  and\ \citenamefont {Zar\'{a}nd}}]{Legeza_DMNRGmanual}%
  \BibitemOpen
  \bibfield  {author} {\bibinfo {author} {\bibfnamefont {\"{O}.}\ \bibnamefont
  {Legeza}}, \bibinfo {author} {\bibfnamefont {C.P.}\ \bibnamefont {Moca}},
  \bibinfo {author} {\bibfnamefont {A.I.}\ \bibnamefont {T\'{o}th}}, \bibinfo
  {author} {\bibfnamefont {I.}~\bibnamefont {Weymann}}, \ and\ \bibinfo
  {author} {\bibfnamefont {G.}~\bibnamefont {Zar\'{a}nd}},\ }\href@noop {}
  {\enquote {\bibinfo {title} {{Manual for the flexible DM-NRG code}},}\
  }\bibinfo {howpublished} {arXiv:0809.3143v1} (\bibinfo {year} {2008}),\
  \bibinfo {note} {(the open access Budapest code is available at
  http://www.phy.bme.hu/\~{}dmnrg/)}\BibitemShut {NoStop}%
\bibitem [{\citenamefont {T\'{o}th}\ \emph {et~al.}(2008)\citenamefont
  {T\'{o}th}, \citenamefont {Moca}, \citenamefont {Legeza},\ and\ \citenamefont
  {Zar\'{a}nd}}]{Toth_Phys.Rev.B78/2008}%
  \BibitemOpen
  \bibfield  {author} {\bibinfo {author} {\bibfnamefont {A.I.}\ \bibnamefont
  {T\'{o}th}}, \bibinfo {author} {\bibfnamefont {C.P.}\ \bibnamefont {Moca}},
  \bibinfo {author} {\bibfnamefont {\"{O}.}\ \bibnamefont {Legeza}}, \ and\
  \bibinfo {author} {\bibfnamefont {G.}~\bibnamefont {Zar\'{a}nd}},\ }\bibfield
   {title} {\enquote {\bibinfo {title} {{Density matrix numerical
  renormalization group for non-Abelian symmetries}},}\ }\href@noop {}
  {\bibfield  {journal} {\bibinfo  {journal} {Phys. Rev. B}\ }\textbf {\bibinfo
  {volume} {78}},\ \bibinfo {pages} {245109} (\bibinfo {year}
  {2008})}\BibitemShut {NoStop}%
\bibitem [{\citenamefont {Oliveira}\ and\ \citenamefont
  {Oliveira}(1994)}]{OliveiraPhysRevB.49.11986}%
  \BibitemOpen
  \bibfield  {author} {\bibinfo {author} {\bibfnamefont {W.~C.}\ \bibnamefont
  {Oliveira}}\ and\ \bibinfo {author} {\bibfnamefont {L.~N.}\ \bibnamefont
  {Oliveira}},\ }\bibfield  {title} {\enquote {\bibinfo {title} {Generalized
  numerical renormalization-group method to calculate the thermodynamical
  properties of impurities in metals},}\ }\href@noop {} {\bibfield  {journal}
  {\bibinfo  {journal} {Phys. Rev. B}\ }\textbf {\bibinfo {volume} {49}},\
  \bibinfo {pages} {11986} (\bibinfo {year} {1994})}\BibitemShut {NoStop}%
\bibitem [{\citenamefont {A.~N.~Pasupathy}\ and\ \citenamefont
  {Ralph}(2004)}]{Pasupathy_Sc.306/86}%
  \BibitemOpen
  \bibfield  {author} {\bibinfo {author} {\bibfnamefont {J.~Martinek J. E.
  Grose L. A. K. Donev P. L. Mc~Euen}\ \bibnamefont {A.~N.~Pasupathy},
  \bibfnamefont {R.~C.~Bialczak}}\ and\ \bibinfo {author} {\bibfnamefont
  {D.~C.}\ \bibnamefont {Ralph}},\ }\bibfield  {title} {\enquote {\bibinfo
  {title} {{The Kondo effect in the presence of ferromagnetism.}}}\ }\href@noop
  {} {\bibfield  {journal} {\bibinfo  {journal} {Science}\ }\textbf {\bibinfo
  {volume} {306}},\ \bibinfo {pages} {86} (\bibinfo {year} {2004})}\BibitemShut
  {NoStop}%
\bibitem [{\citenamefont {H.~B.~Heersche}\ and\ \citenamefont
  {Chuang}(2006)}]{Heersche_PRL.96/017205}%
  \BibitemOpen
  \bibfield  {author} {\bibinfo {author} {\bibfnamefont {A.~Folk L. P.
  Kouwenhoven H. S. J. Zant A. A. Houck J.~Labaziewicz}\ \bibnamefont
  {H.~B.~Heersche}, \bibfnamefont {Z.~Groot}}\ and\ \bibinfo {author}
  {\bibfnamefont {I.~L.}\ \bibnamefont {Chuang}},\ }\bibfield  {title}
  {\enquote {\bibinfo {title} {{Kondo effect in the presence of magnetic
  impurities.}}}\ }\href@noop {} {\bibfield  {journal} {\bibinfo  {journal}
  {Phys. Rev. Lett.}\ }\textbf {\bibinfo {volume} {96}},\ \bibinfo {pages}
  {017205} (\bibinfo {year} {2006})}\BibitemShut {NoStop}%
\bibitem [{\citenamefont {K.~Hamaya}\ and\ \citenamefont
  {Machida}(2008)}]{Hamaya_PRB.77/081302}%
  \BibitemOpen
  \bibfield  {author} {\bibinfo {author} {\bibfnamefont {K.~Shibata M. Jung M.
  Kawamura S. Ishida T. Taniyama K. Hirakawa Y. Arakawa~T.}\ \bibnamefont
  {K.~Hamaya}, \bibfnamefont {M.~Kitabatake}}\ and\ \bibinfo {author}
  {\bibnamefont {Machida}},\ }\bibfield  {title} {\enquote {\bibinfo {title}
  {{Oscillatory changes in the tunneling magnetoresistance effect in
  semiconductor quantum-dot spin valves.}}}\ }\href@noop {} {\bibfield
  {journal} {\bibinfo  {journal} {Phys. Rev. B}\ }\textbf {\bibinfo {volume}
  {77}},\ \bibinfo {pages} {081302(R)} (\bibinfo {year} {2008})}\BibitemShut
  {NoStop}%
\bibitem [{\citenamefont {J.~R.~Hauptmann}\ and\ \citenamefont
  {Lindelof}(2008)}]{Hauptmann_PRB.4/373-376}%
  \BibitemOpen
  \bibfield  {author} {\bibinfo {author} {\bibfnamefont {J.~Paaske}\
  \bibnamefont {J.~R.~Hauptmann}}\ and\ \bibinfo {author} {\bibfnamefont
  {P.~E.}\ \bibnamefont {Lindelof}},\ }\bibfield  {title} {\enquote {\bibinfo
  {title} {{Electric-field-controlled spin reversal in a quantum dot with
  ferromagnetic contacts.}}}\ }\href@noop {} {\bibfield  {journal} {\bibinfo
  {journal} {Nature Physics}\ }\textbf {\bibinfo {volume} {4}},\ \bibinfo
  {pages} {373 -- 376} (\bibinfo {year} {2008})}\BibitemShut {NoStop}%
\bibitem [{\citenamefont {Gaass}\ \emph {et~al.}(2011)\citenamefont {Gaass},
  \citenamefont {H\"uttel}, \citenamefont {Kang}, \citenamefont {Weymann},
  \citenamefont {von Delft},\ and\ \citenamefont
  {Strunk}}]{Gaass_Phys.Rev.Lett.107/2011}%
  \BibitemOpen
  \bibfield  {author} {\bibinfo {author} {\bibfnamefont {M.}~\bibnamefont
  {Gaass}}, \bibinfo {author} {\bibfnamefont {A.~K.}\ \bibnamefont {H\"uttel}},
  \bibinfo {author} {\bibfnamefont {K.}~\bibnamefont {Kang}}, \bibinfo {author}
  {\bibfnamefont {I.}~\bibnamefont {Weymann}}, \bibinfo {author} {\bibfnamefont
  {J.}~\bibnamefont {von Delft}}, \ and\ \bibinfo {author} {\bibfnamefont
  {Ch.}\ \bibnamefont {Strunk}},\ }\bibfield  {title} {\enquote {\bibinfo
  {title} {{Universality of the Kondo effect in quantum dots with ferromagnetic
  leads}},}\ }\href@noop {} {\bibfield  {journal} {\bibinfo  {journal} {Phys.
  Rev. Lett.}\ }\textbf {\bibinfo {volume} {107}},\ \bibinfo {pages} {176808}
  (\bibinfo {year} {2011})}\BibitemShut {NoStop}%
\bibitem [{\citenamefont {Martinek}\ \emph
  {et~al.}(2003{\natexlab{b}})\citenamefont {Martinek}, \citenamefont {Sindel},
  \citenamefont {Borda}, \citenamefont {Barna\'{s}}, \citenamefont {K\"onig},
  \citenamefont {Sch\"on},\ and\ \citenamefont {von
  Delft}}]{Martinek_Phys.Rev.Lett.91/2003_247202}%
  \BibitemOpen
  \bibfield  {author} {\bibinfo {author} {\bibfnamefont {J.}~\bibnamefont
  {Martinek}}, \bibinfo {author} {\bibfnamefont {M.}~\bibnamefont {Sindel}},
  \bibinfo {author} {\bibfnamefont {L.}~\bibnamefont {Borda}}, \bibinfo
  {author} {\bibfnamefont {J.}~\bibnamefont {Barna\'{s}}}, \bibinfo {author}
  {\bibfnamefont {J.}~\bibnamefont {K\"onig}}, \bibinfo {author} {\bibfnamefont
  {G.}~\bibnamefont {Sch\"on}}, \ and\ \bibinfo {author} {\bibfnamefont
  {J.}~\bibnamefont {von Delft}},\ }\bibfield  {title} {\enquote {\bibinfo
  {title} {{Kondo effect in the presence of itinerant-electron ferromagnetism
  studied with the numerical renormalization group method}},}\ }\href@noop {}
  {\bibfield  {journal} {\bibinfo  {journal} {Phys. Rev. Lett.}\ }\textbf
  {\bibinfo {volume} {91}},\ \bibinfo {pages} {247202} (\bibinfo {year}
  {2003}{\natexlab{b}})}\BibitemShut {NoStop}%
\bibitem [{\citenamefont {\'{S}wirkowicz}\ \emph {et~al.}(2006)\citenamefont
  {\'{S}wirkowicz}, \citenamefont {Wilczy\'{n}ski}, \citenamefont
  {Wawrzyniak},\ and\ \citenamefont
  {Barna\'{s}}}]{Swirkowicz_Phys.Rev.B73/2006}%
  \BibitemOpen
  \bibfield  {author} {\bibinfo {author} {\bibfnamefont {R.}~\bibnamefont
  {\'{S}wirkowicz}}, \bibinfo {author} {\bibfnamefont {M.}~\bibnamefont
  {Wilczy\'{n}ski}}, \bibinfo {author} {\bibfnamefont {M.}~\bibnamefont
  {Wawrzyniak}}, \ and\ \bibinfo {author} {\bibfnamefont {J.}~\bibnamefont
  {Barna\'{s}}},\ }\bibfield  {title} {\enquote {\bibinfo {title} {{Kondo
  effect in quantum dots coupled to ferromagnetic leads with noncollinear
  magnetizations}},}\ }\href@noop {} {\bibfield  {journal} {\bibinfo  {journal}
  {Phys. Rev. B}\ }\textbf {\bibinfo {volume} {73}},\ \bibinfo {pages} {193312}
  (\bibinfo {year} {2006})}\BibitemShut {NoStop}%
\bibitem [{\citenamefont {Weymann}(2011)}]{Weymann_Phys.Rev.B83/2011}%
  \BibitemOpen
  \bibfield  {author} {\bibinfo {author} {\bibfnamefont {I.}~\bibnamefont
  {Weymann}},\ }\bibfield  {title} {\enquote {\bibinfo {title}
  {Finite-temperature spintronic transport through kondo quantum dots:
  Numerical renormalization group study},}\ }\href@noop {} {\bibfield
  {journal} {\bibinfo  {journal} {Phys. Rev. B}\ }\textbf {\bibinfo {volume}
  {83}},\ \bibinfo {pages} {113306} (\bibinfo {year} {2011})}\BibitemShut
  {NoStop}%
\bibitem [{\citenamefont {Sindel}\ \emph {et~al.}(2007)\citenamefont {Sindel},
  \citenamefont {Borda}, \citenamefont {Martinek}, \citenamefont {Bulla},
  \citenamefont {K\"{o}nig}, \citenamefont {Sch\"{o}n}, \citenamefont
  {Maekawa},\ and\ \citenamefont {von Delft}}]{Sindel_Phys.Rev.B76/2007}%
  \BibitemOpen
  \bibfield  {author} {\bibinfo {author} {\bibfnamefont {M.}~\bibnamefont
  {Sindel}}, \bibinfo {author} {\bibfnamefont {L.}~\bibnamefont {Borda}},
  \bibinfo {author} {\bibfnamefont {J.}~\bibnamefont {Martinek}}, \bibinfo
  {author} {\bibfnamefont {R.}~\bibnamefont {Bulla}}, \bibinfo {author}
  {\bibfnamefont {J.}~\bibnamefont {K\"{o}nig}}, \bibinfo {author}
  {\bibfnamefont {G.}~\bibnamefont {Sch\"{o}n}}, \bibinfo {author}
  {\bibfnamefont {S.}~\bibnamefont {Maekawa}}, \ and\ \bibinfo {author}
  {\bibfnamefont {J.}~\bibnamefont {von Delft}},\ }\bibfield  {title} {\enquote
  {\bibinfo {title} {{Kondo quantum dot coupled to ferromagnetic leads:
  Numerical renormalization group study}},}\ }\href@noop {} {\bibfield
  {journal} {\bibinfo  {journal} {Phys. Rev. B}\ }\textbf {\bibinfo {volume}
  {76}},\ \bibinfo {pages} {45321} (\bibinfo {year} {2007})}\BibitemShut
  {NoStop}%
\bibitem [{\citenamefont {Misiorny}\ and\ \citenamefont
  {Barna\'{s}}(2009)}]{Misiorny_Phys.Stat.Sol.B246/2009}%
  \BibitemOpen
  \bibfield  {author} {\bibinfo {author} {\bibfnamefont {M.}~\bibnamefont
  {Misiorny}}\ and\ \bibinfo {author} {\bibfnamefont {J.}~\bibnamefont
  {Barna\'{s}}},\ }\bibfield  {title} {\enquote {\bibinfo {title} {Switching of
  molecular magnets},}\ }\href@noop {} {\bibfield  {journal} {\bibinfo
  {journal} {Phys. Stat. Sol. B}\ }\textbf {\bibinfo {volume} {246}},\ \bibinfo
  {pages} {695} (\bibinfo {year} {2009})}\BibitemShut {NoStop}%
\bibitem [{\citenamefont {K{\"o}nig}\ \emph {et~al.}(2005)\citenamefont
  {K{\"o}nig}, \citenamefont {Martinek}, \citenamefont {Barna{\'s}},\ and\
  \citenamefont {Sch{\"o}n}}]{Konig_Lect.NotesPhys.658/2005}%
  \BibitemOpen
  \bibfield  {author} {\bibinfo {author} {\bibfnamefont {J.}~\bibnamefont
  {K{\"o}nig}}, \bibinfo {author} {\bibfnamefont {J.}~\bibnamefont {Martinek}},
  \bibinfo {author} {\bibfnamefont {J.}~\bibnamefont {Barna{\'s}}}, \ and\
  \bibinfo {author} {\bibfnamefont {G.}~\bibnamefont {Sch{\"o}n}},\ }\bibfield
  {title} {\enquote {\bibinfo {title} {{Quantum dots attached to ferromagnetic
  leads: Exchange field, spin precession, and Kondo effect}},}\ }\href@noop {}
  {\bibfield  {journal} {\bibinfo  {journal} {Lect. Notes Phys.}\ }\textbf
  {\bibinfo {volume} {658}},\ \bibinfo {pages} {146--164} (\bibinfo {year}
  {2005})}\BibitemShut {NoStop}%
\bibitem [{\citenamefont {Martinek}\ \emph {et~al.}(2005)\citenamefont
  {Martinek}, \citenamefont {Sindel}, \citenamefont {Borda}, \citenamefont
  {Barna\ifmmode~\acute{s}\else \'{s}\fi{}}, \citenamefont {Bulla},
  \citenamefont {K\"onig}, \citenamefont {Sch\"on}, \citenamefont {Maekawa},\
  and\ \citenamefont {von Delft}}]{Martinek_Phys.Rev.B72/2005}%
  \BibitemOpen
  \bibfield  {author} {\bibinfo {author} {\bibfnamefont {J.}~\bibnamefont
  {Martinek}}, \bibinfo {author} {\bibfnamefont {M.}~\bibnamefont {Sindel}},
  \bibinfo {author} {\bibfnamefont {L.}~\bibnamefont {Borda}}, \bibinfo
  {author} {\bibfnamefont {J.}~\bibnamefont {Barna\ifmmode~\acute{s}\else
  \'{s}\fi{}}}, \bibinfo {author} {\bibfnamefont {R.}~\bibnamefont {Bulla}},
  \bibinfo {author} {\bibfnamefont {J.}~\bibnamefont {K\"onig}}, \bibinfo
  {author} {\bibfnamefont {G.}~\bibnamefont {Sch\"on}}, \bibinfo {author}
  {\bibfnamefont {S.}~\bibnamefont {Maekawa}}, \ and\ \bibinfo {author}
  {\bibfnamefont {J.}~\bibnamefont {von Delft}},\ }\bibfield  {title} {\enquote
  {\bibinfo {title} {{Gate-controlled spin splitting in quantum dots with
  ferromagnetic leads in the Kondo regime}},}\ }\href@noop {} {\bibfield
  {journal} {\bibinfo  {journal} {Phys. Rev. B}\ }\textbf {\bibinfo {volume}
  {72}},\ \bibinfo {pages} {121302} (\bibinfo {year} {2005})}\BibitemShut
  {NoStop}%
\end{thebibliography}

%

\end{document}